\documentclass[traditabstract]{aa}

\usepackage{graphicx}

\usepackage{txfonts}

\begin{document}

\title{The mass-metallicity relation of SDSS quasars}

\author{K. Matsuoka \inst{1} \and T. Nagao \inst{1,2,3} \and A. Marconi \inst{4} \and R. Maiolino \inst{5} \and Y. Taniguchi \inst{2}}

\institute{Graduate School of Science and Engineering, Ehime University, 2-5 Bunkyo-cho, Matsuyama 790-8577, Japan\\ \email{kenta@cosmos.phys.sci.ehime-u.ac.jp} \and Research Center for Space and Cosmic Evolution, Ehime University, 2-5 Bunkyo-cho, Matsuyama 790-8577, Japan \and Optical and Infrared Astronomy Division, National Astronomical Observatory of Japan, 2-21-1 Osawa, Mitaka 181-8588, Japan \and Dipartimento di Fisica e Astronomia, Universita degli Studi di Firenze, Largo E. Fermi 2, 50125 Firenze, Italy \and INAF -- Osservatorio Astrofisico di Roma, Via di Frascati 33, 00040 Monte Porzio Catone, Italy\\}

\abstract{
Active galactic nuclei (AGNs) are characterized by a clear correlation between luminosity and metallicity ($L_{\rm AGN}$-$Z_{\rm AGN}$ relation). The origin of this correlation is not clear. It may result from a relation between the black hole mass ($M_{\rm BH}$) and metallicity, or from a relation between the accretion rate ($L/L_{\rm Edd}$) and metallicity. To investigate the origin of the $L_{\rm AGN}$-$Z_{\rm AGN}$ relation, we use optical spectra of 2383 quasars at $2.3 < z < 3.0$ from the Sloan Digital Sky Survey. By using this data set, we have constructed composite spectra of 33 subsamples in intervals of both $M_{\rm BH}$ and $L/L_{\rm Edd}$. From these composite spectra we measure emission-line flux ratios that are sensitive to the metallicity of the broad line region (BLR); specifically, \ion{N}{v}$\lambda$1240/\ion{C}{iv}$\lambda$1549, \ion{N}{v}$\lambda$1240/\ion{He}{ii}$\lambda$1640, (\ion{Si}{iv}$\lambda$1398+\ion{O}{iv}]$\lambda$1402)/\ion{C}{iv}$\lambda$1549, and \ion{Al}{iii}$\lambda$1857/\ion{C}{iv}$\lambda$1549. We find that there is a significant correlation between $M_{\rm BH}$ and $Z_{\rm BLR}$ as inferred from all four metallicity-sensitive emission-line flux ratios. This result strongly suggests that the observed $L_{\rm AGN}$-$Z_{\rm AGN}$ relation is mostly a consequence of the $M_{\rm BH}$-$Z_{\rm AGN}$ relation. The relation between $M_{\rm BH}$ and $Z_{\rm BLR}$ is likely a consequence of both the $M_{\rm BH}$-$M_{\rm bul}$ relation and of the mass-metallicity relation in the host galaxy. We also find that $L/L_{\rm Edd}$ correlates with the emission line flux ratios involving \ion{N}{v} (more specifically, \ion{N}{v}/\ion{C}{iv} and \ion{N}{v}/\ion{He}{ii}), while it does not correlate with the other two metallicity sensitive emission line flux ratios, i.e., (\ion{Si}{iv}+\ion{O}{iv}])/\ion{C}{iv} and \ion{Al}{iii}/\ion{C}{iv}. These correlations indicate that the emission-line flux ratios involving \ion{N}{v} depend on both metallicity and relative abundance of nitrogen. We suggest that the relation between $L/L_{\rm Edd}$ and those line ratios involving nitrogen, is caused by a delay of the black hole accretion rate relative to the onset of nuclear star formation of about 10$^8$ years, which is the timescale required for the nitrogen enrichment.
}

\keywords{galaxies: active -- galaxies: evolution -- galaxies: nuclei -- quasars: emission lines -- quasars: general}

\maketitle


\section{Introduction}

In the local universe, there is a tight correlation between the mass of black holes at the center of galaxies ($M_{\rm BH}$) and their hosting stellar spheroids ($M_{\rm bul}$), as well as with their velocity dispersion (e.g., Kormendy \& Richstone 1995; Richstone et al. 1998; Ferrarese \& Merritt 2000; Gebhardt et al. 2000; Tremaine et al. 2002; Marconi \& Hunt 2003; H\"{a}ring \& Rix 2004; Barth et al. 2005). These correlations suggest that the formation and evolution of supermassive black holes (SMBHs) are connected with those of their host galaxies. However the mechanism responsible for such co-evolution is not yet clear. In this study, we approach this problem by focusing on the properties of the active galactic nuclei (AGNs), in particular $M_{\rm BH}$ and accretion rate $L/L_{\rm Edd}$, and their gas metallicity ($Z_{\rm AGN}$).

Hamann \& Ferland (1993) proposed that emission-line flux ratios \ion{N}{v}$\lambda$1240/\ion{C}{iv}$\lambda$1549 and \ion{N}{v}$\lambda$1240/\ion{He}{ii}$\lambda$1640 trace the broad-line region (BLR) metallicity ($Z_{\rm BLR}$) across a wide range of the AGN luminosities ($L_{\rm AGN}$) and BLR densities. They also reported a clear relation between the AGN luminosity and BLR metallicity: more luminous AGNs tend to have more metal-rich BLR clouds (see also Nagao et al. 2006a; Juarez et al. 2009). The AGN luminosity is determined by both the black hole mass and accretion rate relative to its Eddington limit. On the other hand, the metallicity is an important tracer for the formation and evolution of the host galaxy, since it is closely related to the past star forming history of the host galaxy (e.g., Matteucci \& Padovani 1993; Hamann \& Ferland 1993, 1999; Venkatesan et al. 2004). Therefore, it is possible to investigate the formation and evolution of SMBHs and of their host galaxies by comparing $L_{\rm AGN}$ and the gas metallicity. However it is not clear how the gas metallicity of BLR is connected to the chemical properties of their host galaxies. The BLR is a very small region in galactic nuclei ($r_{\rm BLR} < 1$ pc; e.g., Suganuma et al. 2006), which may have evolved more rapidly than their host galaxies. However, a clear relation between the metallicity of narrow-line region (NLR) and the AGN luminosity has been also reported in recent studies (e.g., Nagao et al. 2006b; Matsuoka et al. 2009). In contrast to the BLR, the scale size of the NLR is comparable to the size of the host galaxies ($r_{\rm NLR} \sim 10^{2-4}$ pc; e.g., Bennert et al. 2006a, 2006b) and its gas mass is about $M_{\rm NLR} \sim 10^{5-7} M_\odot$ (much larger than for the BLR, $M_{\rm BLR} \sim 10^{2-4} M_\odot$; see, e.g., Baldwin et al. 2003), implying that the NLR metallicity traces the enrichment in host galaxy. As a consequence, the finding of a $L_{\rm AGN}$-$Z_{\rm NLR}$ relation suggests that also the $L_{\rm AGN}$-$Z_{\rm BLR}$ relation traces a connection between the nuclear engine and the metallicity in the host galaxy, or at least in the central region.

Since in flux-limited samples $M_{\rm BH}$ and $L_{\rm AGN}$ are correlated, Hamann \& Ferland (1993) suggested that the $L_{\rm AGN}$-$Z_{\rm BLR}$ relation is a consequence of a relation between black hole mass and BLR metallicity. However, the $L_{\rm AGN}$-$Z_{\rm BLR}$ relation may also originate from a connection between the gas metallicity and the accretion rate $L/L_{\rm Edd}$. Warner et al. (2004) analyzed rest-frame ultraviolet spectra of type 1 AGNs at $0 < z < 5$ and found that the emission-line flux ratios involving nitrogen lines (expected to trace the BLR metallicity; see e.g., Hamann et al. 2002) show a correlation with the black hole mass, while no correlation with the Eddington ratio. On the other hand, Shemmer et al. (2004), by using near-infrared spectroscopic measurements of luminous quasars at $2 < z < 3.5$, found that the BLR metallicity, as inferred by \ion{N}{v}/\ion{C}{iv}, is primarily correlated with the Eddington ratio and not with the black hole mass. Dietrich et al. (2009) studied 10 high-$z$ quasars and showed that the \ion{N}{v}/\ion{C}{iv} flux ratios indicate super-solar metallicities in the quasars with a high-accretion rate, suggesting a correlation similar to that found by Shemmer et al. (2004), although from those quasars alone there is no correlation between \ion{N}{v}/\ion{C}{iv} and $L/L_{\rm Edd}$ (see their Figure 12). This discrepancy may be attributed to the small size of the adopted samples (578 quasars for Warner et al. 2004 and 92 quasars for Shemmer et al. 2004).

We use a large quasar sample of the Sloan Digital Sky Survey (SDSS; York et al. 2000) to investigate possible correlations between metallicity, black hole mass, and Eddington ratio. We apply a stacking method delivering composite spectra by dividing the quasar sample into several bins of Eddington ratio and of black hole mass. This method allows us to measure emission-line flux ratios unambiguously by using the high signal-to-noise ratio spectra and to break the degeneracy between black hole mass and Eddington ratio in flux limited samples. More specifically, we present the composite spectra for 33 subsamples binned by $M_{\rm BH}$ and $L/L_{\rm Edd}$ out of 2383 quasar spectra obtained from the SDSS, and investigate possible dependence of metallicity-sensitive emission-line flux ratios on $M_{\rm BH}$ and $L/L_{\rm Edd}$. In section 2, we describe the quasar sample and composite spectra. The description of the measurement of emission lines is given in section 3. In section 4 we summarize our results, and in section 5 we discuss the interpretation for the our results. The summary of this paper is given in section 6.


\section{Sample and composite spectra}

We select a large sample of SDSS quasars from the catalog in Shen et al. (2008) on the basis of the following criteria:
\begin{enumerate}
\item
redshift in the range $2.3 < z < 3.0$ to investigate rest-frame ultraviolet spectra including emission lines such as \ion{N}{v}, \ion{C}{iv}, and \ion{He}{ii},
\item
both the luminosity at $1350\AA$ and the line width of \ion{C}{iv} emission are measured, and
\item
identified as non broad absorption line (non-BAL) quasars.
\end{enumerate}
Then we obtain a sample that consists of 2678 SDSS quasars. Their spectroscopic data are available in the SDSS archive, Data Release 5 (Adelman-McCarthy et al. 2007; Schneider et al. 2007). The spectral resolution of the SDSS spectra is $\sim$ 2000. The wavelength coverage is $3800\AA \la \lambda_{\rm obs} \la 9200\AA$, within which we can measure some redshifted ultraviolet emission lines (i.e., \ion{N}{v}$\lambda$1240, \ion{C}{iv}$\lambda$1549, \ion{He}{ii}$\lambda$1640, \ion{Si}{iv}$\lambda$1398+\ion{O}{iv}]$\lambda$1402, and \ion{Al}{iii}$\lambda$1857) to study the metallicity of BLRs.


\begin{figure}[!]
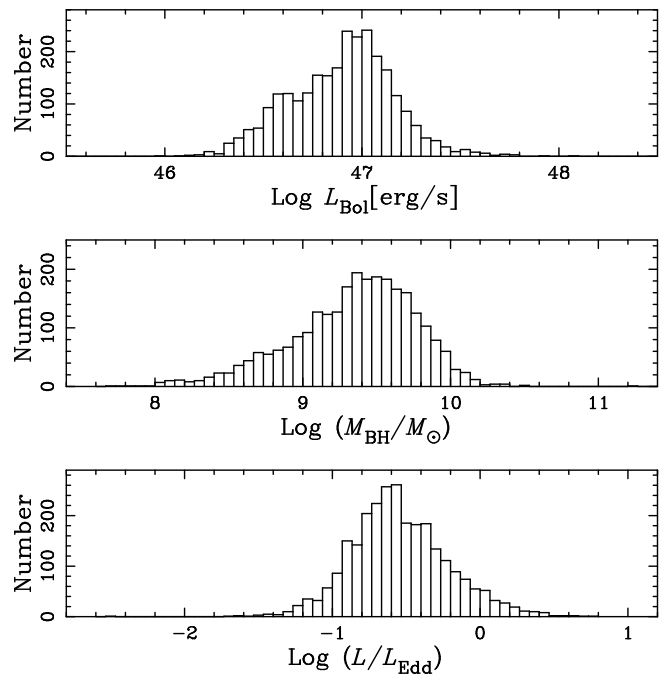

\centering
\begin{tabular}{c}
\includegraphics[width=8.5cm]{figure2010012101.ps}\\
\noalign{\medskip}
\includegraphics[width=8.5cm]{figure2010012102.ps}\\
\noalign{\medskip}
\includegraphics[width=8.5cm]{figure2010012103.ps}\\
\end{tabular}
\caption{Distributions of bolometric luminosity, black hole mass, and the Eddington ratio for our quasar sample that of 2678 SDSS quasars.}
\end{figure}

The black hole mass is estimated by the line width of \ion{C}{iv}$\lambda$1549, adopting the following formula (Vestergaard \& Peterson 2006; Shen et al. 2008):
\begin{equation}
\log \left(\frac{M_{\rm BH}}{M_\odot} \right) = 0.66 + 0.53 \log \left[\frac{\lambda L_\lambda (1350{\rm \AA})}{10^{44} \ {\rm ergs \ s^{-1}}} \right] + 2 \log \left(\frac{\rm FWHM}{{\rm km \ s^{-1}}} \right),
\end{equation}
where FWHM is the full width at half maximum of the \ion{C}{iv} line. We also use the continuum luminosity to estimate the Eddington ratio, $L/L_{\rm Edd}$ with the bolometric correction factor at ${\rm BC} = 3.81$ (Richards et al. 2006; Shen et al. 2008), and we obtain the following formula,
\begin{equation}
\log \left(\frac{L}{L_{\rm Edd}} \right) = 5.82 + 0.47 \log \left[\frac{\lambda L_\lambda (1350{\rm \AA})}{10^{44} \ {\rm ergs \ s^{-1}}} \right] - 2 \log \left(\frac{\rm FWHM}{{\rm km \ s^{-1}}} \right).
\end{equation}
Figure 1 shows the distribution of the luminosity, black hole mass, and Eddington ratio for our SDSS quasar sample. Our sample includes the luminosity at $45.9 < \log L_{\rm bol} ({\rm ergs \ s^{-1}}) < 48.1$, black hole mass at $7.7 < \log (M_{\rm BH}/M_\odot) < 11.3$, and Eddington ratio at $-2.6 < \log (L/L_{\rm Edd}) < 0.8$. Their averages and standard deviations are $46.90\pm0.07$, $9.34\pm0.18$, and $-0.55\pm0.11$, respectively. Note that the low cutoffs in the $L_{\rm bol}$, $M_{\rm BH}$, and $L/L_{\rm Edd}$ distributions presented in Figure 1 are due to the flux-limited nature of the SDSS quasar sample.

To investigate the relationship between the black hole mass and metallicity, independently of the Eddington ratio, or the $L/L_{\rm Edd}$-$Z_{\rm BLR}$ relation independently of $M_{\rm BH}$, we divide our sample into subsamples by fixing one of two parameters of the black hole mass and Eddington ratio. By using these subsamples, we examine possible relationship between the black hole mass and BLR metallicity with a given Eddington ratio, or the $L/L_{\rm Edd}$-$Z_{\rm BLR}$ relation at a given black hole mass. Table 1 gives the detailed information for our subsamples. Note that we exclude some quasars that are at the ranges of the black hole mass, $\log (M_{\rm BH}/M_\odot) < 8.4$ and $10.0 \leq \log (M_{\rm BH}/M_\odot)$, and Eddington ratio, $\log (L/L_{\rm Edd}) < -1.2$ and $0.0 \leq \log (L/L_{\rm Edd})$, and also exclude some subsamples composed with less than 10 quasars in order to improve the statistical significance for results. This implies a final selection of 2383 quasars.

We extract the composite spectra for the 33 subsamples given in Table 1 in the following way. By adopting the redshift defined in the catalog in Shen et al. (2008), we shift the quasar spectra to the rest frame before stacking. After shifting, each spectrum is normalized by using the mean flux density at $1350\AA < \lambda_{\rm rest} < 1365\AA$, where there are no strong emission lines. For the stacking procedure, we use the IRAF task {\tt scombine} and make median spectra (instead of average spectra) for combining pixels with a 3 sigma-clipping rejection. There are two techniques to generate composite spectra ``in general": (1) the median (or mean) spectrum, which preserves the relative fluxes of the emission features; and (2) the ``geometric" mean (i.e., logarithmic mean) spectrum, which preserves the global continuum shape (see, Vanden Berk et al. 2001). Since we are interested in the emission-line flux ratios, we apply the former method (i.e., the linear median spectrum). Figures 2 and 3 show the extracted composite spectra for each bin of black hole mass or Eddington ratio, respectively.


\begin{table}
\caption{Definition of the 33 intervals of $M_{\rm BH}$ and $L/L_{\rm Edd}$ used to create the composite spectra, along with the number of quasars within each interval.}
\begin{tabular}{l l r}
\hline \hline \noalign{\smallskip}
\multicolumn{1}{c}{Black Hole Mass} & \multicolumn{1}{c}{Eddington Ratio} & \multicolumn{1}{c}{$N$}\\
\noalign{\smallskip} \hline \noalign{\smallskip}
$8.4 \leq \log(M_{\rm BH}/M_\odot) < 8.6$ & $-0.4 \leq \log(L/L_{\rm Edd}) < -0.2$ & 19\\
& $-0.2 \leq \log(L/L_{\rm Edd}) < 0.0$ & 32\\
\noalign{\smallskip}
$8.6 \leq \log(M_{\rm BH}/M_\odot) < 8.8$ & $-0.6 \leq \log(L/L_{\rm Edd}) < -0.4$ & 14\\
& $-0.4 \leq \log(L/L_{\rm Edd}) < -0.2$ & 44\\
& $-0.2 \leq \log(L/L_{\rm Edd}) < 0.0$ & 58\\
\noalign{\smallskip}
$8.8 \leq \log(M_{\rm BH}/M_\odot) < 9.0$ & $-0.8 \leq \log(L/L_{\rm Edd}) < -0.6$ & 16\\
& $-0.6 \leq \log(L/L_{\rm Edd}) < -0.4$ & 45\\
& $-0.4 \leq \log(L/L_{\rm Edd}) < -0.2$ & 71\\
& $-0.2 \leq \log(L/L_{\rm Edd}) < 0.0$ & 54\\
\noalign{\smallskip}
$9.0 \leq \log(M_{\rm BH}/M_\odot) < 9.2$ & $-1.0 \leq \log(L/L_{\rm Edd}) < -0.8$ & 13\\
& $-0.8 \leq \log(L/L_{\rm Edd}) < -0.6$ & 71\\
& $-0.6 \leq \log(L/L_{\rm Edd}) < -0.4$ & 87\\
& $-0.4 \leq \log(L/L_{\rm Edd}) < -0.2$ & 108\\
& $-0.2 \leq \log(L/L_{\rm Edd}) < 0.0$ & 55\\
\noalign{\smallskip}
$9.2 \leq \log(M_{\rm BH}/M_\odot) < 9.4$ & $-1.0 \leq \log(L/L_{\rm Edd}) < -0.8$ & 55\\
& $-0.8 \leq \log(L/L_{\rm Edd}) < -0.6$ & 108\\
& $-0.6 \leq \log(L/L_{\rm Edd}) < -0.4$ & 190\\
& $-0.4 \leq \log(L/L_{\rm Edd}) < -0.2$ & 118\\
& $-0.2 \leq \log(L/L_{\rm Edd}) < 0.0$ & 11\\
\noalign{\smallskip}
$9.4 \leq \log(M_{\rm BH}/M_\odot) < 9.6$ & $-1.2 \leq \log(L/L_{\rm Edd}) < -1.0$ & 22\\
& $-1.0 \leq \log(L/L_{\rm Edd}) < -0.8$ & 81\\
& $-0.8 \leq \log(L/L_{\rm Edd}) < -0.6$ & 214\\
& $-0.6 \leq \log(L/L_{\rm Edd}) < -0.4$ & 188\\
& $-0.4 \leq \log(L/L_{\rm Edd}) < -0.2$ & 46\\
\noalign{\smallskip}
$9.6 \leq \log(M_{\rm BH}/M_\odot) < 9.8$ & $-1.2 \leq \log(L/L_{\rm Edd}) < -1.0$ & 25\\
& $-1.0 \leq \log(L/L_{\rm Edd}) < -0.8$ & 126\\
& $-0.8 \leq \log(L/L_{\rm Edd}) < -0.6$ & 196\\
& $-0.6 \leq \log(L/L_{\rm Edd}) < -0.4$ & 74\\
& $-0.4 \leq \log(L/L_{\rm Edd}) < -0.2$ & 18\\
\noalign{\smallskip}
$9.8 \leq \log(M_{\rm BH}/M_\odot) < 10.0$ & $-1.2 \leq \log(L/L_{\rm Edd}) < -1.0$ & 52\\
& $-1.0 \leq \log(L/L_{\rm Edd}) < -0.8$ & 82\\
& $-0.8 \leq \log(L/L_{\rm Edd}) < -0.6$ & 61\\
& $-0.6 \leq \log(L/L_{\rm Edd}) < -0.4$ & 29\\
\noalign{\smallskip} \hline \noalign{\smallskip}
Total & & 2383\\
\noalign{\smallskip} \hline
\end{tabular}
\end{table}


\begin{figure*}[!]
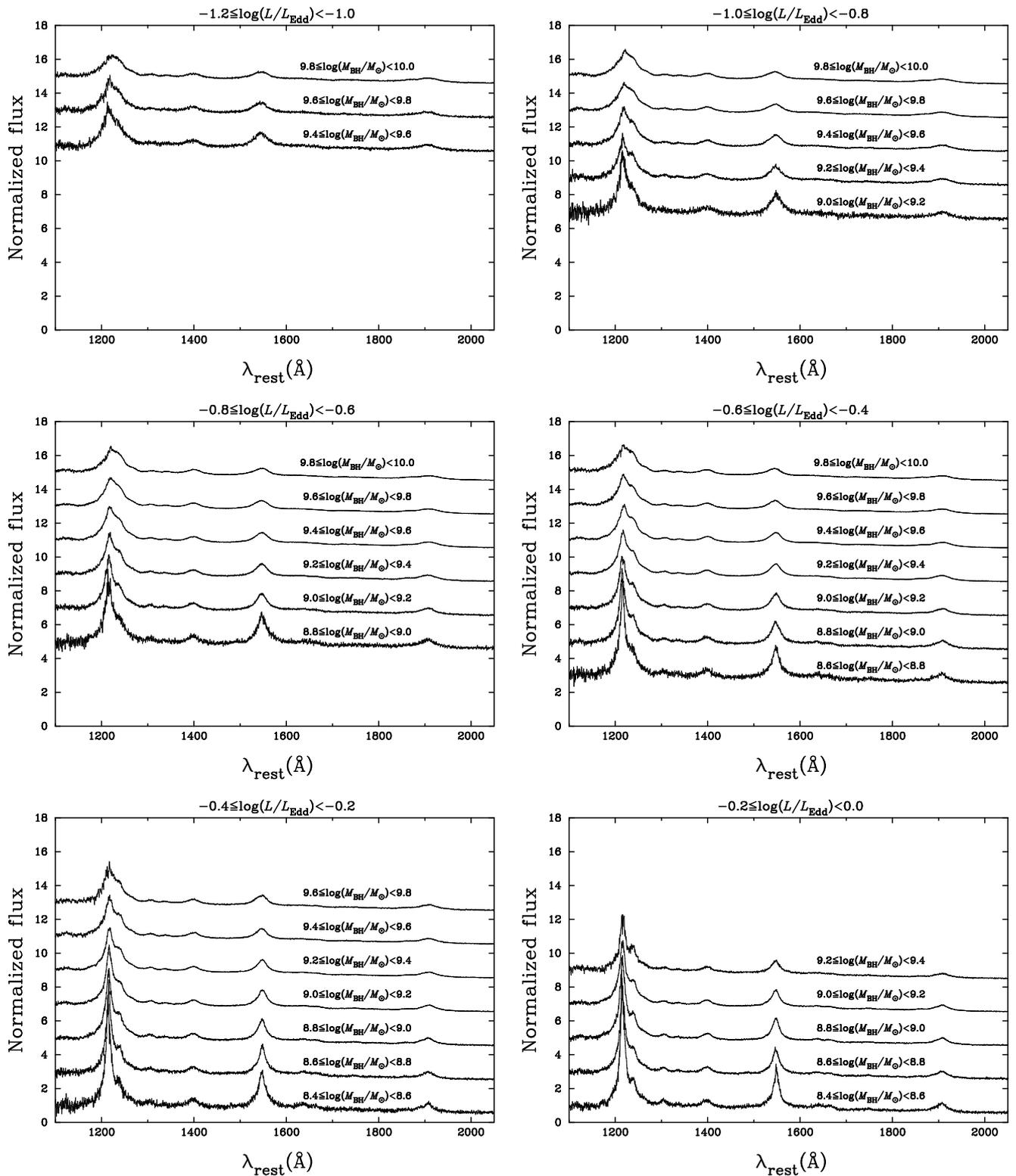

\centering
\begin{tabular}{c c}
\includegraphics[width=8.5cm]{figure2010010201.ps} & \includegraphics[width=8.5cm]{figure2010010202.ps}\\
\noalign{\medskip}
\includegraphics[width=8.5cm]{figure2010010203.ps} & \includegraphics[width=8.5cm]{figure2010010204.ps}\\
\noalign{\medskip}
\includegraphics[width=8.5cm]{figure2010010205.ps} & \includegraphics[width=8.5cm]{figure2010010206.ps}\\
\end{tabular}
\caption{Each panel shows the composite spectra in a fixed interval of Eddington ratio by varying $M_{\rm BH}$, normalized by the mean flux density at $1350\AA < \lambda_{\rm rest} < 1365\AA$. The height scale is the same for all spectra, but different spectra are shifted by two.}
\end{figure*}


\begin{figure*}[!]
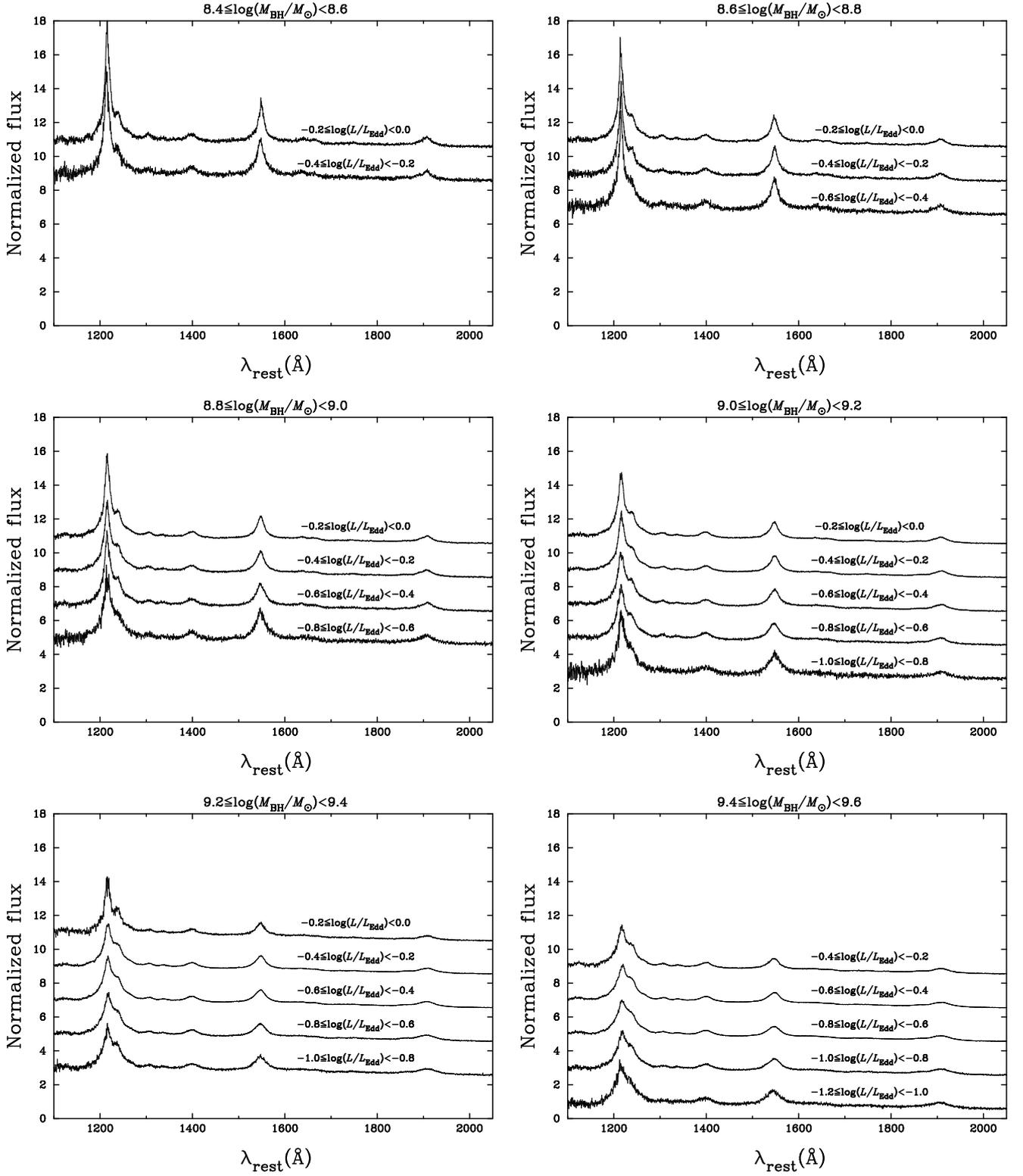

\centering
\begin{tabular}{c c}
\includegraphics[width=8.5cm]{figure2010010207.ps} & \includegraphics[width=8.5cm]{figure2010010208.ps}\\
\noalign{\medskip}
\includegraphics[width=8.5cm]{figure2010010209.ps} & \includegraphics[width=8.5cm]{figure2010010210.ps}\\
\noalign{\medskip}
\includegraphics[width=8.5cm]{figure2010010211.ps} & \includegraphics[width=8.5cm]{figure2010010212.ps}\\
\end{tabular}
\caption{Each panel shows the composite spectra in a fixed interval of $M_{\rm BH}$ by varying the Eddington ratio. The height scale is the same for all spectra, but different spectra are shifted by two.}
\end{figure*}


\setcounter{figure}{2}
\begin{figure*}[!]
\centering
\begin{tabular}{c c}
\includegraphics[width=8.5cm]{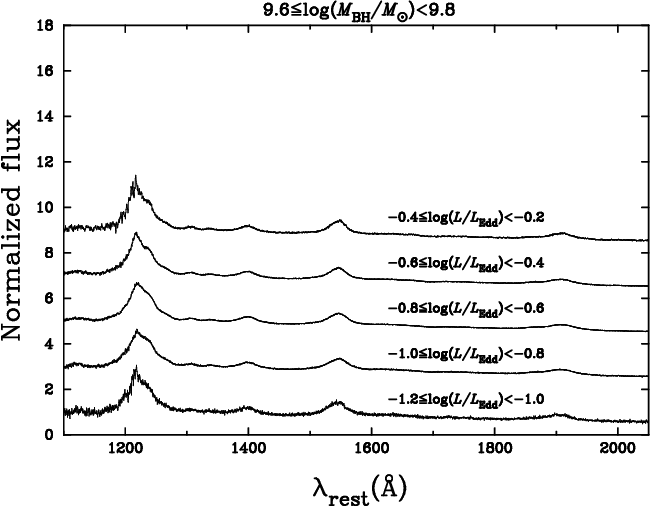} & \includegraphics[width=8.5cm]{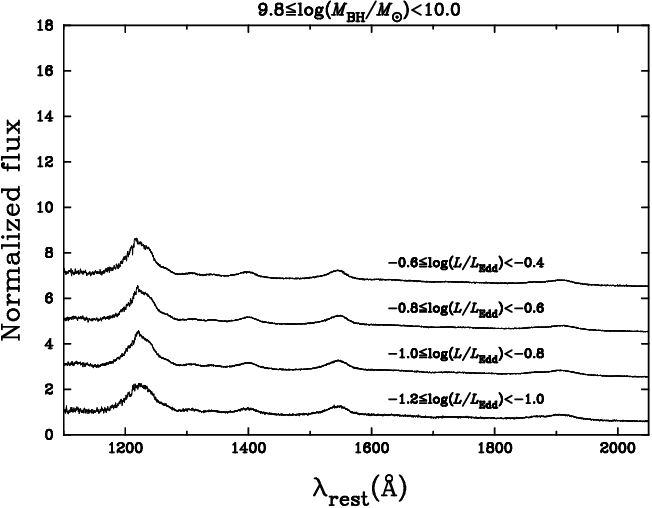}\\
\end{tabular}
\caption{Continued.}
\end{figure*}


\section{Emission-line measurement}

It is sometimes difficult to measure broad emission-line fluxes because some close emission-lines (e.g., Ly$\alpha$ and \ion{N}{v}$\lambda$1240) are heavily blended and because the continuum level is often not easy to estimate due to the contribution of both weak emission lines and \ion{Fe}{ii} multiplet features. Previous studies have mostly used one of the two following methods. One is to fit the detected emission-line features with some appropriate profile function; the other method consists in defining a local continuum level for each emission line and integrating the flux above the adopted continuum level. The former method is far more accurate when the flux measurements requires the deblending of emission lines, which is the case for most quasar UV spectra. We therefore adopt the former method in this study.

We use the following function to fit the line profiles:
\begin{equation}
F_\lambda = \cases{F_0 \times \left(\frac{\lambda}{\lambda_0} \right)^{-\alpha} \ \ \ \ \ \ \ {\rm for} \ \lambda \ge \lambda_0 \cr F_0 \times \left(\frac{\lambda}{\lambda_0} \right)^{+\beta} \ \ \ \ \ \ \ {\rm for} \ \lambda < \lambda_0, \cr}
\end{equation}
where the power-law indices ($\alpha$ and $\beta$) are different between the blue side and red side of a given emission line (i.e., $\alpha \neq \beta$ generally). This function achieves a better fit than double-Gaussian and modified (i.e., asymmetric) Lorentzian methods (see Nagao et al. 2006a for more details). Since the emission-line profile depends on the ionization degree (e.g., Gaskell 1982; Wilkes 1984, 1986; Baskin \& Laor 2005), we divide ultraviolet emission lines into the following two groups: high-ionization lines (HILs; \ion{N}{v}$\lambda$1240, \ion{O}{iv}$\lambda$1402, \ion{N}{iv}]$\lambda$1486, \ion{C}{iv}$\lambda$1549, and \ion{He}{ii}$\lambda$1640) and low-ionization lines (LILs; \ion{Si}{ii}$\lambda$1263, \ion{Si}{iv}$\lambda$1398, \ion{O}{iii}]$\lambda$1663, \ion{Al}{ii}$\lambda$1671, \ion{Al}{iii}$\lambda$1857, \ion{Si}{ii}$\lambda$1887, and \ion{C}{iii}]$\lambda$1909), with the boundary ionization potential of $\sim$ 40 eV (e.g., Collin-Souffrin \& Lasota 1988 for discussion on this dichotomy). We assume that all emission lines in the same group have the same power-law indices of $\alpha$ and $\beta$, and the same wavelength shift of the emission line center ($\lambda_0$) from the systemic velocity. The line profiles for the same group are fit simultaneously to infer the best values of $\alpha$, $\beta$, and $\lambda_0$. The absorption by intervening intergalactic matter on the blue side of Ly$\alpha$ affects the profile fitting for Ly$\alpha$. Therefore, for the Ly$\alpha$ fitting, we adopt the red-side index $\alpha$ inferred for HILs, and the other two parameters ($\beta$ and $\lambda_0$) are left to be free. The model fitting is performed in the spectral region at $1210\AA < \lambda_{\rm rest} < 2000\AA$. The slope and the amplitude of the power-law continuum are initially estimated in two wavelength regions where the emission line contribution to the total flux appears to be small ($1445\AA < \lambda_{\rm rest} < 1455\AA$ and $1973\AA < \lambda_{\rm rest} < 1983\AA$) and finally determined by the model fitting. The fitting results are shown in Figure 4. The best fit power-law indices, $\alpha$ and $\beta$, and the systemic velocity of each line are given in Table 2.

The \ion{O}{i}+\ion{Si}{ii} composite at 1305$\AA$ and \ion{C}{ii}$\lambda$1335 are measured simply by summing up all of fluxes above the continuum level for each line ($1286\AA < \lambda_{\rm rest} < 1322\AA$ for the \ion{O}{i}+\ion{Si}{ii} composite and $1322\AA < \lambda_{\rm rest} < 1357\AA$), because their velocity profiles are different from both HILs and LILs. In each spectrum we detect a ``1600$\AA$ bump". This unidentified emission feature has been noted in earlier studies, e.g., Wilkes (1984), Boyle (1990), Laor et al. (1994), and Nagao et al. (2006a). The possible nature of this feature has been discussed in Nagao et al. (2006a) and it will not be discussed further in this paper. The spectral regions discussed above (\ion{O}{i}+\ion{Si}{ii}$\lambda$1305, \ion{C}{ii}$\lambda$1335, and 1600$\AA$ bump) are excluded from our fitting procedure. The wavelength region at $1687\AA < \lambda_{\rm rest} < 1833\AA$ is also excluded because it includes heavily blended emission lines such as \ion{N}{iv}$\lambda$1719, \ion{Al}{ii}$\lambda$1722, \ion{N}{iii}]$\lambda$1750, and \ion{Fe}{ii} multiplets, which are very difficult to deblend. \ion{N}{iii}]$\lambda$1750 would also be useful to estimate the metallicity in addition to \ion{N}{v}$\lambda$1240 (Shields et al. 1976; Baldwin \& Netzer 1978; Hamann et al. 2002; Dietrich et al. 2003), but we did not take it into account because of the difficulty in deblending it from broad \ion{Fe}{ii} emission.


\section{Results}

Based on our fitting results described in section 3, we derive the emission-line flux ratios that are sensitive to the BLR metallicity, i.e., \ion{N}{v}/\ion{C}{iv}, \ion{N}{v}/\ion{He}{ii}, (\ion{Si}{iv}+\ion{O}{iv}])/\ion{C}{iv}, and \ion{Al}{iii}/\ion{C}{iv} (Hamann et al. 2002; Dietrich et al. 2003; Nagao et al. 2006a; see also Juarez et al. 2009). Table 3 lists the inferred emission-line flux ratios. The error on the emission-line flux ratios is estimated by taking account both statistical and systematic errors. The systematic error (about five percent of the line flux) is associated with the uncertainties in the line-profile fitting.

To investigate the $M_{\rm BH}$-$Z_{\rm BLR}$ relation and $L/L_{\rm Edd}$-$Z_{\rm BLR}$ relation, we show the dependence of the emission-line flux ratios on $M_{\rm BH}$ and $L/L_{\rm Edd}$ in Figure 5. The left-hand panels in Figure 5 show that there is a tight correlation between the metallicity-sensitive emission-line flux ratios and $M_{\rm BH}$. On the other hand, the right-hand panels show that there is a large scatter and no clear dependence of the emission-line flux ratios on the Eddington ratio. These results are consistent with the earlier findings of Warner et al. (2004), who analyzed rest-frame ultraviolet spectra of type 1 AGNs and found the correlation between $M_{\rm BH}$ and $Z_{\rm BLR}$. In contrast, these results appear to be inconsistent with the results of Shemmer et al. (2004), who found a significant relation between $L/L_{\rm Edd}$ and the \ion{N}{v}/\ion{C}{iv} line flux ratio while the dispersion in the relation between $M_{\rm BH}$ and \ion{N}{v}/\ion{C}{iv} is much larger (see also Dietrich et al. 2009). It should be noted that the sample of Shemmer et al. (2004) includes AGNs both at high-$z$ and in the local universe, and the large dispersion in the relation between $M_{\rm BH}$ and \ion{N}{v}/\ion{C}{iv} in their sample is mainly caused by local narrow-line Seyfert 1 galaxies (NLS1s). However, it is known that NLS1s do not follow usual trends in terms of BLR emission-line flux ratios (Shemmer \& Netzer 2002), that is also shown by Warner et al. (2004). Actually, the dispersion in the $M_{\rm BH}$-\ion{N}{v}/\ion{C}{iv} relation in the sample of Shemmer et al. (2004) is significantly reduced if the NLS1s are removed. Note that most of the high-$M_{\rm BH}$ objects in the sample of Shemmer et al. (2004) are very luminous quasars and with high-$L/L_{\rm Edd}$, because their high-$z$ sample was observed spectroscopically in the near-infrared and therefore high-luminosity objects were preferentially selected. Therefore, due to the resulting strong degeneracy between $M_{\rm BH}$ and $L/L_{\rm Edd}$, the \ion{N}{v}/\ion{C}{iv} flux ratio in the sample of Shemmer et al. (2004) shows significant dependence on both $M_{\rm BH}$ and $L/L_{\rm Edd}$.

In order to further investigate the dependence of $Z_{\rm BLR}$ on $M_{\rm BH}$ and $L/L_{\rm Edd}$, it is crucial to examine the $M_{\rm BH}$-$Z_{\rm BLR}$ relation and the $L/L_{\rm Edd}$-$Z_{\rm BLR}$ relation {\it independently}. For instance, if we plot the \ion{N}{v}/\ion{C}{iv} flux ratio as a function of $L/L_{\rm Edd}$ regardless of $M_{\rm BH}$, possible correlations between \ion{N}{v}/\ion{C}{iv} and $M_{\rm BH}$ would prevent us from investigating the relation between \ion{N}{v}/\ion{C}{iv} and $L/L_{\rm Edd}$. In order to minimize the effect of $L/L_{\rm Edd}$ or $M_{\rm BH}$ on any relation between the emission-line flux ratios and $M_{\rm BH}$ or $L/L_{\rm Edd}$ respectively, we normalize the emission-line flux ratios by the values measured from the composite spectra at $-0.6 \leq \log (L/L_{\rm Edd}) < -0.4$ or $9.2 \leq \log (M_{\rm BH}/M_\odot) < 9.4$, respectively. In Figure 6, we show the normalized flux ratios as a function of $M_{\rm BH}$ or $L/L_{\rm Edd}$. In the left-hand panels of Figure 6, we find that there are clear relations between $M_{\rm BH}$ and emission-line flux ratios as seen in Figure 5. On the other hand, surprisingly, we also find that there is a correlation between $L/L_{\rm Edd}$ and the emission-line flux ratios involving \ion{N}{v} in the right panels, which was not obvious in Figure 5. To assess these relations quantitatively, we apply the Spearman rank-order test to statistically examine the possible correlations between $Z_{\rm BLR}$ and $M_{\rm BH}$, or $L/L_{\rm Edd}$. We calculate the Spearman rank-order correlation coefficient $r_{\rm s}$ and their statistical significance $p(r_{\rm s})$, which are given in Table 4. The results of the Spearman rank-order test are summarized as follows:
\begin{itemize}
\item
there are significant correlations between the black hole mass and all four emission-line ratios, and
\item
the line flux ratios involving \ion{N}{v} (i.e., \ion{N}{v}/\ion{C}{iv} and \ion{N}{v}/\ion{He}{ii}) are positively correlated with the Eddington ratio, while the other two flux ratios, i.e., (\ion{Si}{iv}+\ion{O}{iv}])/\ion{C}{iv} and \ion{Al}{iii}/\ion{C}{iv}, do not show any significant correlation with $L/L_{\rm Edd}$.
\end{itemize}


\begin{figure*}[!]
\centering
\begin{tabular}{c c}
\includegraphics[width=8.5cm]{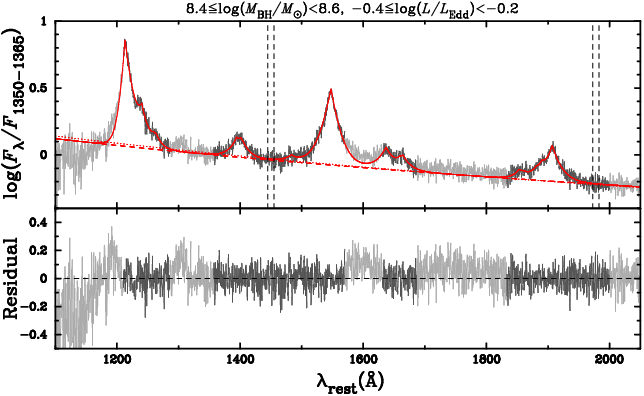} & \includegraphics[width=8.5cm]{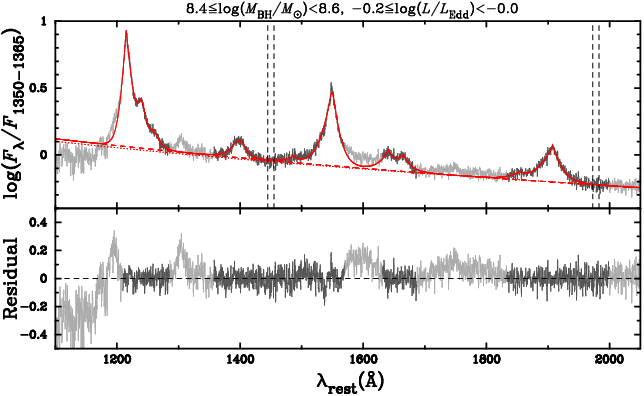}\\
\noalign{\medskip}
\includegraphics[width=8.5cm]{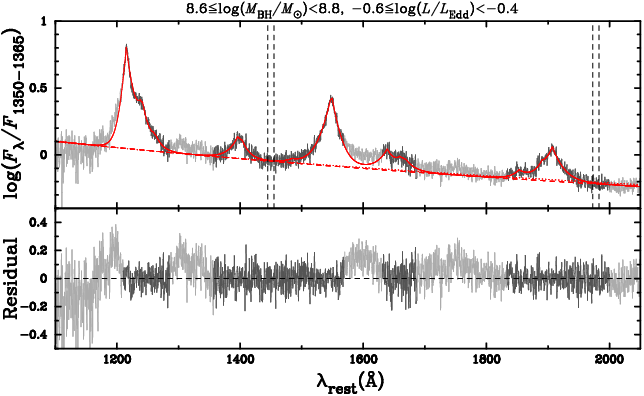} & \includegraphics[width=8.5cm]{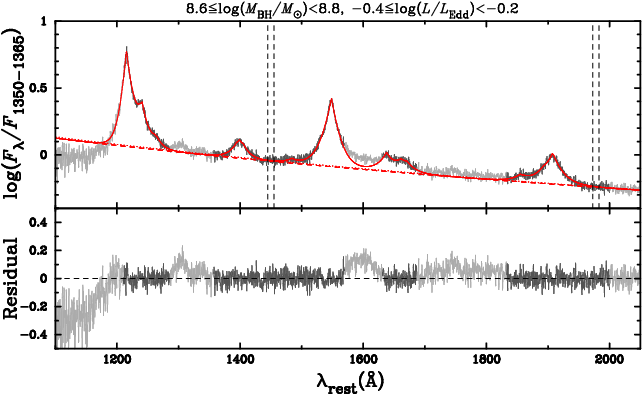}\\
\noalign{\medskip}
\includegraphics[width=8.5cm]{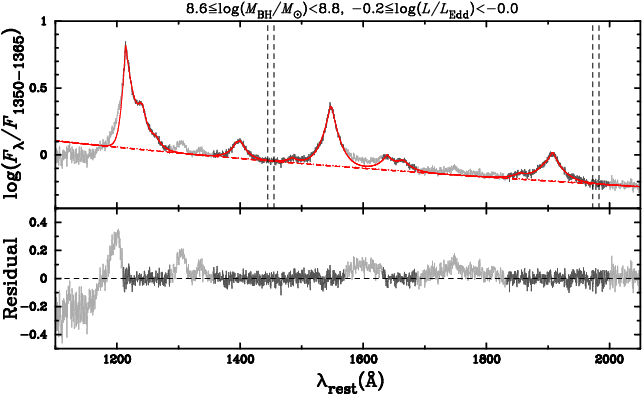} & \includegraphics[width=8.5cm]{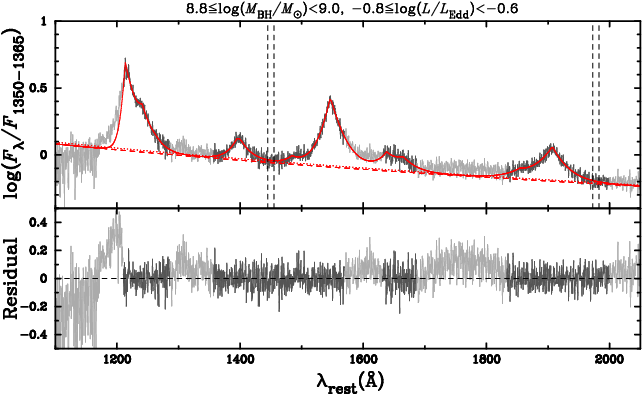}\\
\noalign{\medskip}
\includegraphics[width=8.5cm]{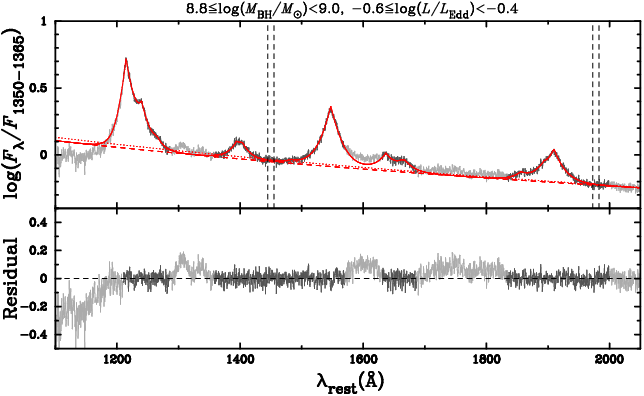} & \includegraphics[width=8.5cm]{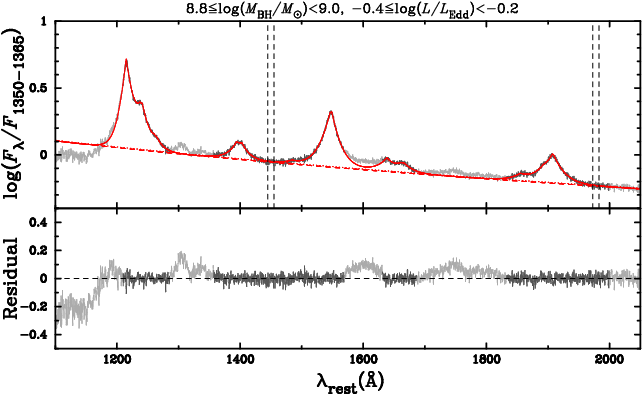}\\
\end{tabular}
\caption{Composite spectra of each quasar subsample and fitting results, shown in the logarithmic scale (upper panels). The residuals are shown in the lower panels. Dark grey lines show the wavelength regions used for the fitting, while the wavelength regions excluded from the fitting process are shown in light grey. The spectral regions denoted by vertical dashed lines are used to determine the initial guess of the continuum level and slope in the fitting process. The initial guess and the fitted power-law continuum are denoted by red dotted and dashed lines, respectively.}
\end{figure*}


\setcounter{figure}{3}
\begin{figure*}[!]
\centering
\begin{tabular}{c c}
\includegraphics[width=8.5cm]{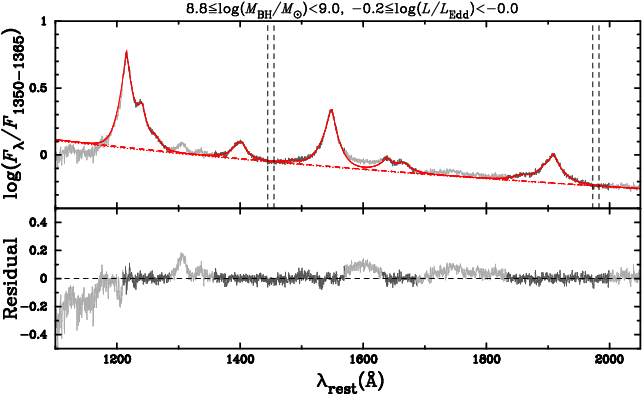} & \includegraphics[width=8.5cm]{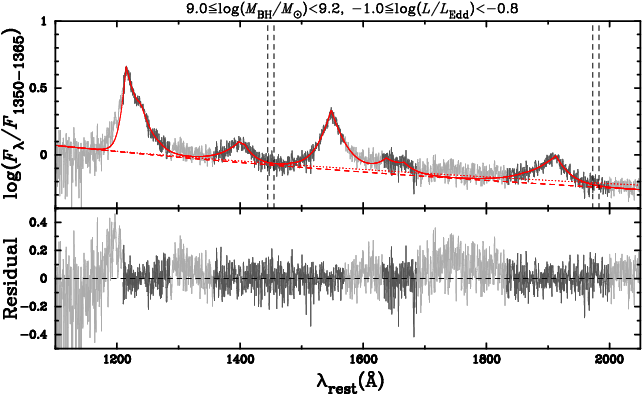}\\
\noalign{\bigskip}\noalign{\bigskip}
\includegraphics[width=8.5cm]{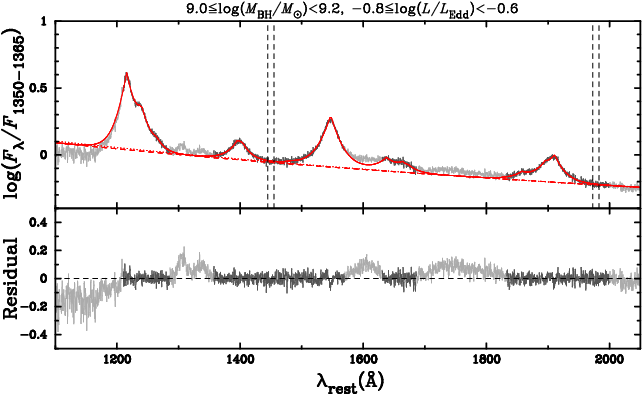} & \includegraphics[width=8.5cm]{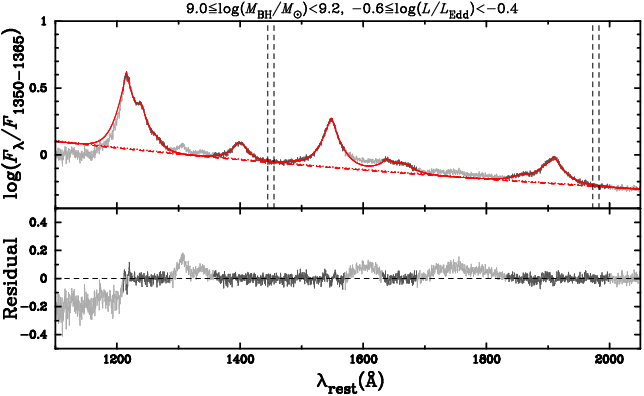}\\
\noalign{\bigskip}\noalign{\bigskip}
\includegraphics[width=8.5cm]{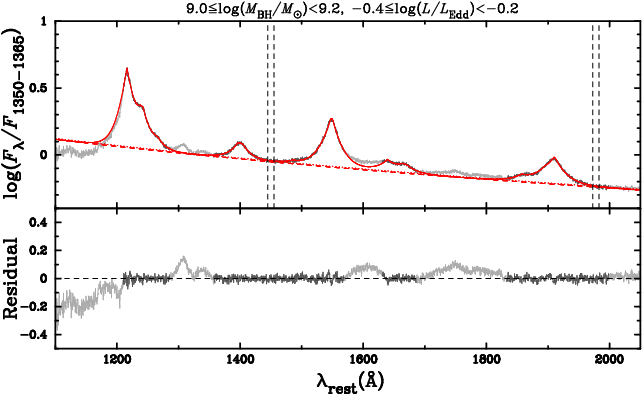} & \includegraphics[width=8.5cm]{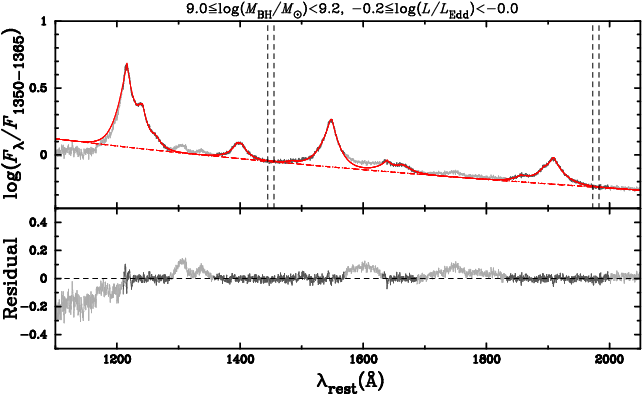}\\
\noalign{\bigskip}\noalign{\bigskip}
\includegraphics[width=8.5cm]{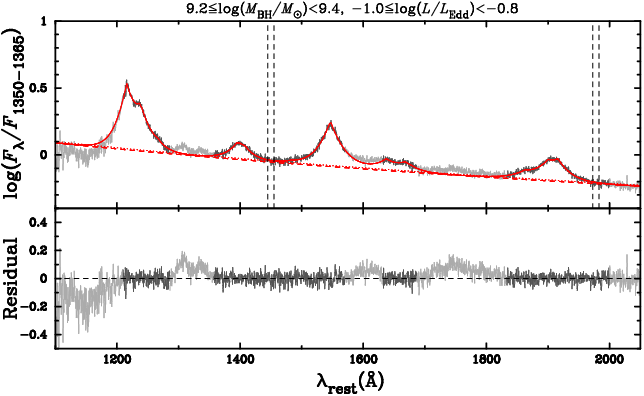} & \includegraphics[width=8.5cm]{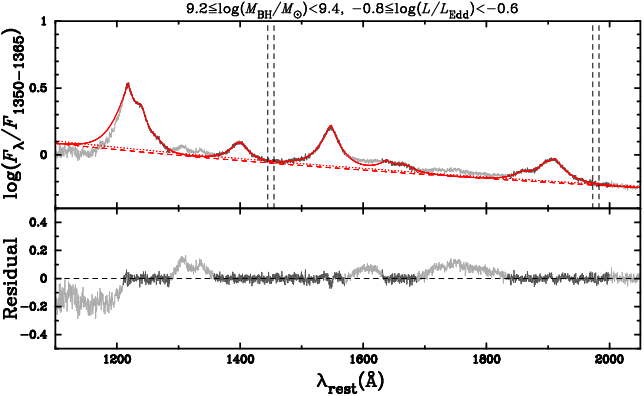}\\
\end{tabular}
\caption{Continued.}
\end{figure*}


\setcounter{figure}{3}
\begin{figure*}[!]
\centering
\begin{tabular}{c c}
\includegraphics[width=8.5cm]{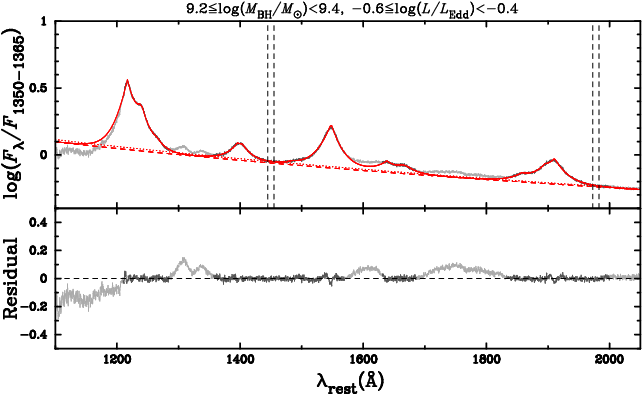} & \includegraphics[width=8.5cm]{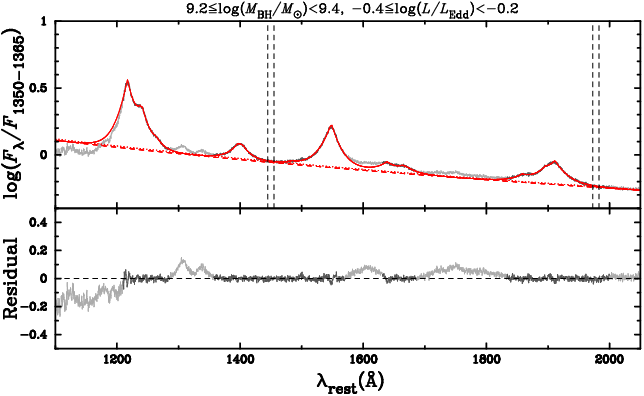}\\
\noalign{\bigskip}\noalign{\bigskip}
\includegraphics[width=8.5cm]{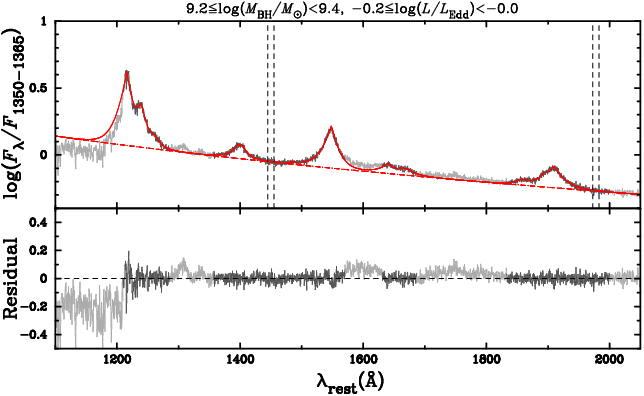} & \includegraphics[width=8.5cm]{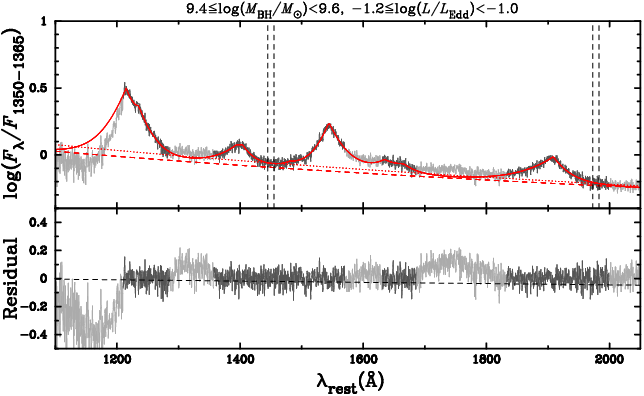}\\
\noalign{\bigskip}\noalign{\bigskip}
\includegraphics[width=8.5cm]{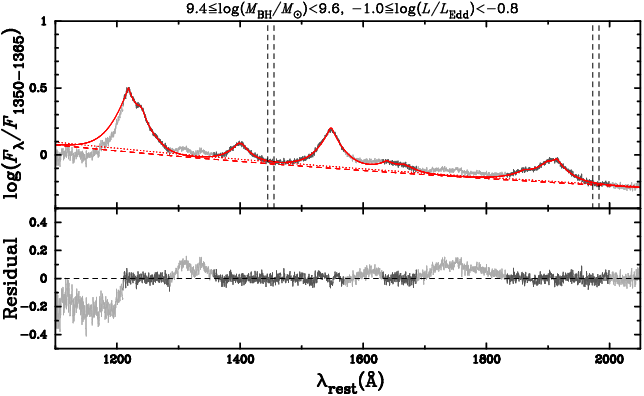} & \includegraphics[width=8.5cm]{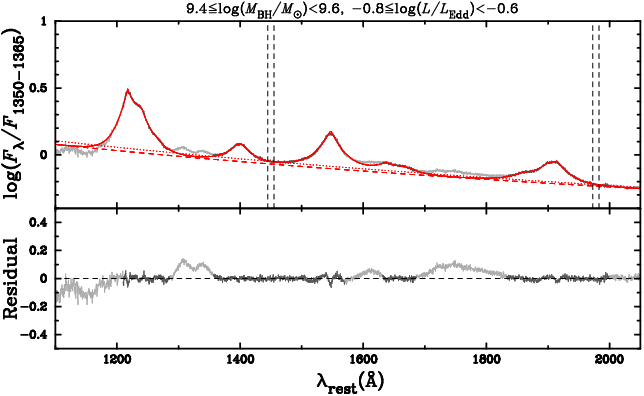}\\
\noalign{\bigskip}\noalign{\bigskip}
\includegraphics[width=8.5cm]{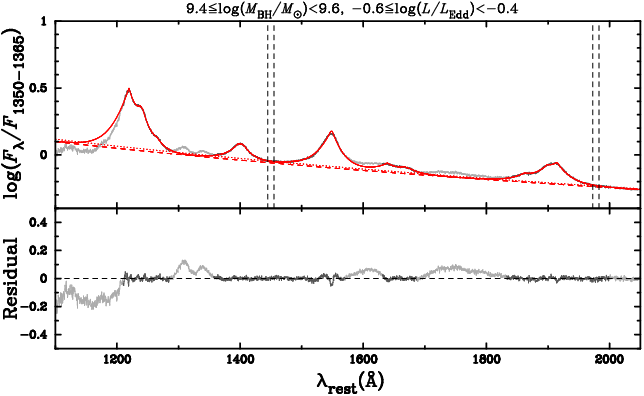} & \includegraphics[width=8.5cm]{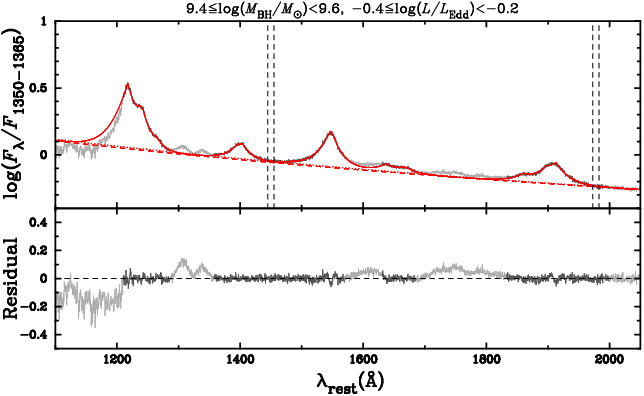}\\
\end{tabular}
\caption{Continued.}
\end{figure*}


\setcounter{figure}{3}
\begin{figure*}[!]
\centering
\begin{tabular}{c c}
\includegraphics[width=8.5cm]{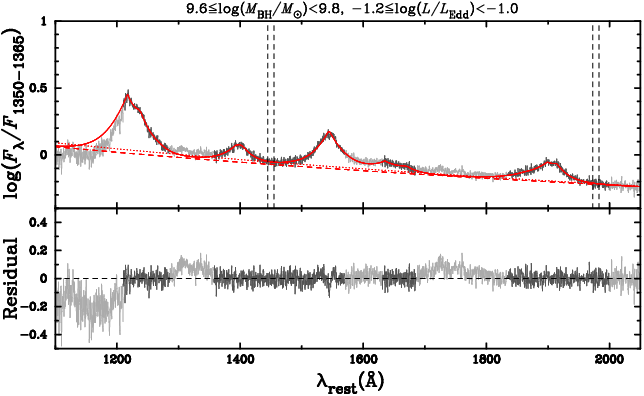} & \includegraphics[width=8.5cm]{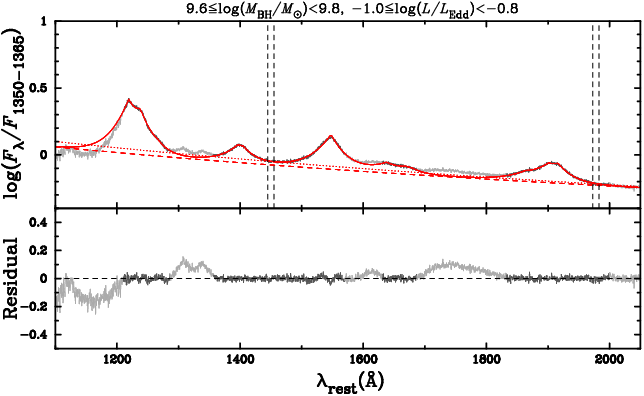}\\
\noalign{\bigskip}\noalign{\bigskip}
\includegraphics[width=8.5cm]{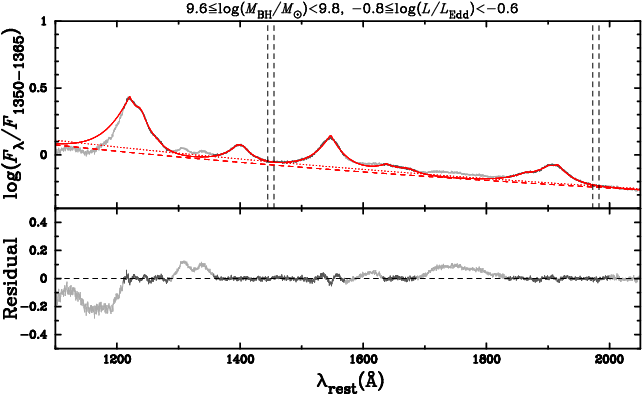} & \includegraphics[width=8.5cm]{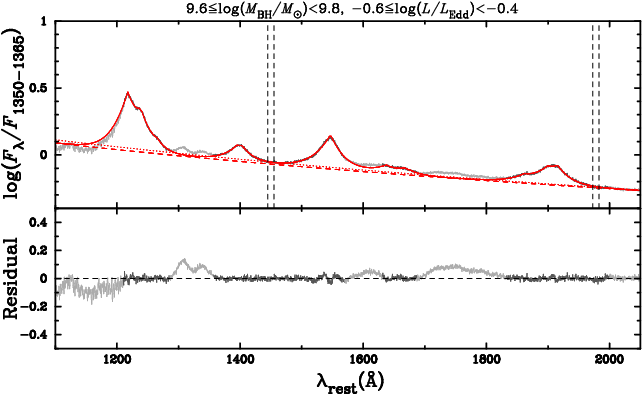}\\
\noalign{\bigskip}\noalign{\bigskip}
\includegraphics[width=8.5cm]{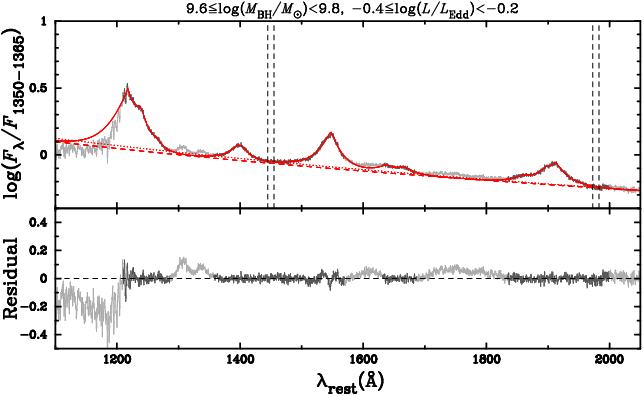} & \includegraphics[width=8.5cm]{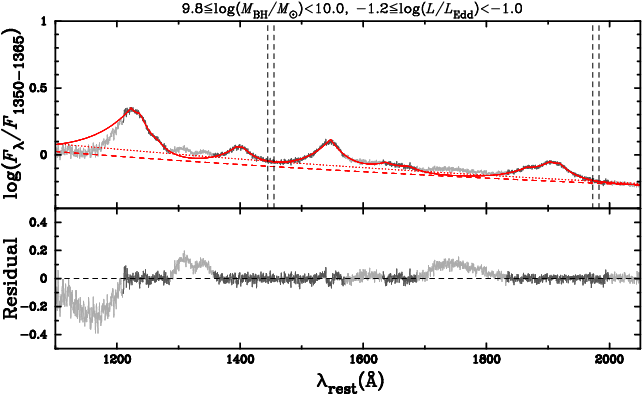}\\
\noalign{\bigskip}\noalign{\bigskip}
\includegraphics[width=8.5cm]{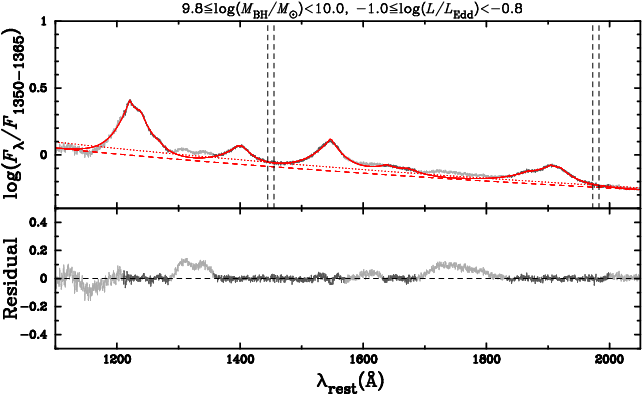} & \includegraphics[width=8.5cm]{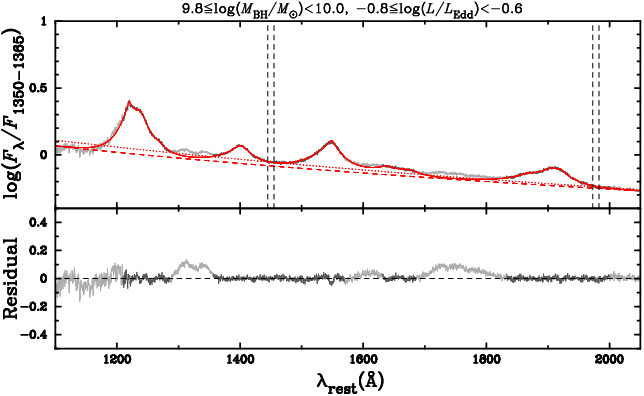}\\
\end{tabular}
\caption{Continued.}
\end{figure*}


\setcounter{figure}{3}
\begin{figure*}[!]
\centering
\begin{tabular}{c}
\includegraphics[width=8.5cm]{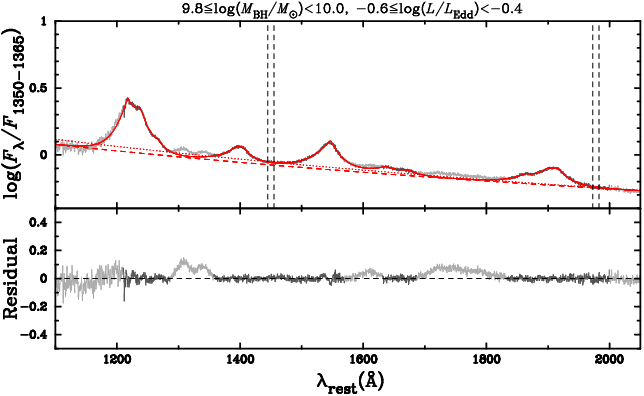}
\end{tabular}
\caption{Continued.}
\end{figure*}


\begin{table*}
\caption{The line profile and the velocity shift of HILs and LILs.}
\begin{tabular}{l l c r r r}
\hline \hline \noalign{\smallskip}
\multicolumn{1}{c}{Black Hole Mass} & \multicolumn{1}{c}{Eddington Ratio} & \multicolumn{1}{c}{Line} & \multicolumn{2}{c}{Line Profile} & \multicolumn{1}{c}{Velocity Shift $^a$}\\
\noalign{\smallskip}
& & & \multicolumn{1}{c}{$\alpha$} & \multicolumn{1}{c}{$\beta$} & \multicolumn{1}{c}{[km/s]}\\
\noalign{\smallskip} \hline \noalign{\smallskip}
$8.4 \leq \log(M_{\rm BH}/M_\odot) < 8.6$  & $-0.4 \leq \log(L/L_{\rm Edd}) < -0.2$ & HIL & $123.02\pm0.05$ & $110.88\pm0.04$ & $ -605.6\pm0.4$\\
& & LIL & $138.66\pm0.11$ & $178.94\pm0.36$ & $ -281.4\pm0.5$\\
& $-0.2 \leq \log(L/L_{\rm Edd}) < 0.0$ & HIL & $165.74\pm0.06$ & $123.38\pm0.04$ & $   64.5\pm0.5$\\
& & LIL & $141.34\pm0.11$ & $110.46\pm0.18$ & $ -159.2\pm0.5$\\
\noalign{\smallskip}
$8.6 \leq \log(M_{\rm BH}/M_\odot) < 8.8$  & $-0.6 \leq \log(L/L_{\rm Edd}) < -0.4$ & HIL & $127.87\pm0.06$ & $101.77\pm0.04$ & $ -266.4\pm0.6$\\
& & LIL & $118.05\pm0.10$ & $140.65\pm0.29$ & $ -248.8\pm0.6$\\
& $-0.4 \leq \log(L/L_{\rm Edd}) < -0.2$ & HIL & $132.01\pm0.06$ & $116.01\pm0.04$ & $ -455.8\pm0.3$\\
& & LIL & $122.14\pm0.12$ & $120.84\pm0.27$ & $   81.5\pm1.3$\\
& $-0.2 \leq \log(L/L_{\rm Edd}) < 0.0$ & HIL & $129.78\pm0.06$ & $122.62\pm0.06$ & $ -287.8\pm0.6$\\
& & LIL & $117.38\pm0.12$ & $109.67\pm0.22$ & $   -3.5\pm1.3$\\
\noalign{\smallskip}
$8.8 \leq \log(M_{\rm BH}/M_\odot) < 9.0$  & $-0.8 \leq \log(L/L_{\rm Edd}) < -0.6$ & HIL & $ 81.28\pm0.05$ & $101.93\pm0.04$ & $ -444.2\pm0.8$\\
& & LIL & $ 81.38\pm0.08$ & $ 69.78\pm0.11$ & $  -96.5\pm2.7$\\
& $-0.6 \leq \log(L/L_{\rm Edd}) < -0.4$ & HIL & $103.79\pm0.05$ & $101.96\pm0.05$ & $ -668.9\pm0.3$\\
& & LIL & $119.18\pm0.11$ & $123.08\pm0.34$ & $  227.4\pm1.2$\\
& $-0.4 \leq \log(L/L_{\rm Edd}) < -0.2$ & HIL & $121.21\pm0.06$ & $104.21\pm0.05$ & $ -381.9\pm0.7$\\
& & LIL & $116.67\pm0.12$ & $106.54\pm0.21$ & $   97.4\pm1.4$\\
& $-0.2 \leq \log(L/L_{\rm Edd}) < 0.0$ & HIL & $142.79\pm0.07$ & $113.59\pm0.05$ & $ -276.5\pm0.7$\\
& & LIL & $124.61\pm0.12$ & $106.89\pm0.25$ & $  147.0\pm1.5$\\
\noalign{\smallskip}
$9.0 \leq \log(M_{\rm BH}/M_\odot) < 9.2$  & $-1.0 \leq \log(L/L_{\rm Edd}) < -0.8$ & HIL & $ 77.59\pm0.05$ & $ 85.55\pm0.05$ & $ -520.0\pm0.9$\\
& & LIL & $ 90.39\pm0.10$ & $ 49.41\pm0.07$ & $  988.6\pm0.9$\\
& $-0.8 \leq \log(L/L_{\rm Edd}) < -0.6$ & HIL & $ 95.05\pm0.06$ & $ 86.71\pm0.05$ & $ -470.5\pm1.0$\\
& & LIL & $121.39\pm0.14$ & $ 96.99\pm0.19$ & $  548.1\pm1.3$\\
& $-0.6 \leq \log(L/L_{\rm Edd}) < -0.4$ & HIL & $ 98.13\pm0.06$ & $ 92.36\pm0.05$ & $ -316.5\pm1.0$\\
& & LIL & $102.07\pm0.11$ & $ 87.34\pm0.16$ & $  391.4\pm1.4$\\
& $-0.4 \leq \log(L/L_{\rm Edd}) < -0.2$ & HIL & $108.65\pm0.07$ & $ 95.04\pm0.05$ & $ -169.0\pm1.0$\\
& & LIL & $105.07\pm0.12$ & $ 93.23\pm0.18$ & $  472.7\pm1.6$\\
& $-0.2 \leq \log(L/L_{\rm Edd}) < 0.0$ & HIL & $129.39\pm0.07$ & $100.76\pm0.06$ & $ -315.6\pm0.8$\\
& & LIL & $111.97\pm0.12$ & $120.67\pm0.31$ & $  148.6\pm1.5$\\
\noalign{\smallskip}
$9.2 \leq \log(M_{\rm BH}/M_\odot) < 9.4$  & $-1.0 \leq \log(L/L_{\rm Edd}) < -0.8$ & HIL & $ 81.43\pm0.06$ & $ 83.96\pm0.06$ & $ -586.6\pm1.2$\\
& & LIL & $ 98.97\pm0.14$ & $ 77.96\pm0.16$ & $ 1313.4\pm1.4$\\
& $-0.8 \leq \log(L/L_{\rm Edd}) < -0.6$ & HIL & $ 88.26\pm0.07$ & $ 75.74\pm0.05$ & $ -483.7\pm1.2$\\
& & LIL & $ 88.12\pm0.12$ & $ 91.19\pm0.21$ & $  605.9\pm1.5$\\
& $-0.6 \leq \log(L/L_{\rm Edd}) < -0.4$ & HIL & $ 91.99\pm0.07$ & $ 79.27\pm0.05$ & $ -377.1\pm1.1$\\
& & LIL & $ 94.80\pm0.12$ & $ 86.50\pm0.21$ & $  504.4\pm1.8$\\
& $-0.4 \leq \log(L/L_{\rm Edd}) < -0.2$ & HIL & $101.37\pm0.08$ & $ 86.08\pm0.06$ & $ -326.4\pm1.1$\\
& & LIL & $ 93.55\pm0.13$ & $ 94.11\pm0.24$ & $  569.3\pm1.8$\\
& $-0.2 \leq \log(L/L_{\rm Edd}) < 0.0$ & HIL & $134.51\pm0.09$ & $100.59\pm0.06$ & $  -36.7\pm0.7$\\
& & LIL & $109.86\pm0.15$ & $144.88\pm0.41$ & $  227.6\pm1.2$\\
\noalign{\smallskip}
$9.4 \leq \log(M_{\rm BH}/M_\odot) < 9.6$  & $-1.2 \leq \log(L/L_{\rm Edd}) < -1.0$ & HIL & $ 61.10\pm0.05$ & $ 71.61\pm0.05$ & $-1066.5\pm1.7$\\
& & LIL & $ 73.61\pm0.10$ & $ 37.91\pm0.04$ & $  218.0\pm2.3$\\
& $-1.0 \leq \log(L/L_{\rm Edd}) < -0.8$ & HIL & $ 73.84\pm0.07$ & $ 69.71\pm0.05$ & $ -367.1\pm1.6$\\
& & LIL & $ 85.60\pm0.13$ & $ 63.20\pm0.16$ & $ 1040.9\pm2.0$\\
& $-0.8 \leq \log(L/L_{\rm Edd}) < -0.6$ & HIL & $ 72.02\pm0.07$ & $ 66.98\pm0.06$ & $ -503.1\pm1.6$\\
& & LIL & $ 84.97\pm0.13$ & $ 63.83\pm0.18$ & $ 1031.0\pm2.3$\\
& $-0.6 \leq \log(L/L_{\rm Edd}) < -0.4$ & HIL & $ 95.55\pm0.09$ & $ 71.92\pm0.06$ & $ -195.7\pm1.4$\\
& & LIL & $ 95.18\pm0.15$ & $ 80.44\pm0.20$ & $ 1108.0\pm1.8$\\
& $-0.4 \leq \log(L/L_{\rm Edd}) < -0.2$ & HIL & $101.12\pm0.08$ & $ 76.95\pm0.06$ & $ -354.3\pm1.3$\\
& & LIL & $102.99\pm0.15$ & $ 92.29\pm0.22$ & $  889.8\pm1.7$\\
\noalign{\smallskip}
$9.6 \leq \log(M_{\rm BH}/M_\odot) < 9.8$  & $-1.2 \leq \log(L/L_{\rm Edd}) < -1.0$ & HIL & $ 65.87\pm0.07$ & $ 64.40\pm0.06$ & $ -975.1\pm1.9$\\
& & LIL & $100.74\pm0.19$ & $ 51.78\pm0.17$ & $ 1120.5\pm2.2$\\
& $-1.0 \leq \log(L/L_{\rm Edd}) < -0.8$ & HIL & $ 67.35\pm0.08$ & $ 52.80\pm0.05$ & $ -467.1\pm1.8$\\
& & LIL & $ 83.13\pm0.16$ & $ 53.26\pm0.16$ & $ 1415.7\pm2.4$\\
& $-0.8 \leq \log(L/L_{\rm Edd}) < -0.6$ & HIL & $ 70.16\pm0.08$ & $ 58.12\pm0.06$ & $ -491.7\pm1.8$\\
& & LIL & $ 83.86\pm0.16$ & $ 55.64\pm0.16$ & $ 1384.7\pm2.4$\\
& $-0.6 \leq \log(L/L_{\rm Edd}) < -0.4$ & HIL & $ 81.89\pm0.08$ & $ 64.19\pm0.06$ & $ -538.3\pm1.7$\\
& & LIL & $ 93.55\pm0.16$ & $ 68.60\pm0.19$ & $ 1184.9\pm1.8$\\
& $-0.4 \leq \log(L/L_{\rm Edd}) < -0.2$ & HIL & $ 92.17\pm0.08$ & $ 70.41\pm0.06$ & $ -528.2\pm1.3$\\
& & LIL & $ 94.59\pm0.14$ & $ 86.38\pm0.23$ & $  773.8\pm1.8$\\
\noalign{\smallskip}
$9.8 \leq \log(M_{\rm BH}/M_\odot) < 10.0$ & $-1.2 \leq \log(L/L_{\rm Edd}) < -1.0$ & HIL & $ 60.67\pm0.09$ & $ 44.19\pm0.06$ & $ -680.0\pm1.9$\\
& & LIL & $ 75.89\pm0.16$ & $ 48.30\pm0.14$ & $ 1723.7\pm2.2$\\
& $-1.0 \leq \log(L/L_{\rm Edd}) < -0.8$ & HIL & $ 65.97\pm0.09$ & $ 48.31\pm0.06$ & $ -396.6\pm2.1$\\
& & LIL & $ 73.31\pm0.17$ & $ 43.66\pm0.12$ & $ 1995.4\pm2.2$\\
& $-0.8 \leq \log(L/L_{\rm Edd}) < -0.6$ & HIL & $ 73.09\pm0.09$ & $ 47.22\pm0.06$ & $ -208.3\pm1.9$\\
& & LIL & $ 79.10\pm0.18$ & $ 46.35\pm0.15$ & $ 2018.2\pm2.6$\\
& $-0.6 \leq \log(L/L_{\rm Edd}) < -0.4$ & HIL & $ 82.84\pm0.09$ & $ 51.92\pm0.06$ & $ -585.6\pm1.6$\\
& & LIL & $ 97.84\pm0.19$ & $ 63.54\pm0.20$ & $ 1480.2\pm1.9$\\
\noalign{\smallskip} \hline
\end{tabular}

\begin{list}{}{}
\item[$^a$]
The shift of the emission line center ($\lambda_0$) from the systemic velocity.
\end{list}

\end{table*}


\begin{table*}
\caption{The measured line flux ratios.}
\begin{tabular}{l l c c c c}
\hline \hline \noalign{\smallskip}
\multicolumn{1}{c}{Black Hole Mass} & \multicolumn{1}{c}{Eddington Ratio} & \multicolumn{4}{c}{Line Flux Ratios $^a$}\\
\noalign{\smallskip}
& & \multicolumn{1}{c}{\ion{N}{v}/\ion{C}{iv}} & \multicolumn{1}{c}{\ion{N}{v}/\ion{He}{ii}} & \multicolumn{1}{c}{(\ion{Si}{iv}+\ion{O}{iv}])/\ion{C}{iv}} & \multicolumn{1}{c}{\ion{Al}{iii}/\ion{C}{iv}}\\
\noalign{\smallskip} \hline \noalign{\smallskip}
$8.4 \leq \log(M_{\rm BH}/M_\odot) < 8.6$ & $-0.4 \leq \log(L/L_{\rm Edd}) < -0.2$ & $0.320\pm0.024$ & $1.974\pm0.144$ & $0.197\pm0.015$ & $0.049\pm0.004$\\
& $-0.2 \leq \log(L/L_{\rm Edd}) < 0.0$ & $0.460\pm0.034$ & $3.556\pm0.259$ & $0.193\pm0.013$ & $0.045\pm0.003$\\
\noalign{\smallskip}
$8.6 \leq \log(M_{\rm BH}/M_\odot) < 8.8$ & $-0.6 \leq \log(L/L_{\rm Edd}) < -0.4$ & $0.489\pm0.037$ & $2.900\pm0.212$ & $0.237\pm0.018$ & $0.055\pm0.004$\\
& $-0.4 \leq \log(L/L_{\rm Edd}) < -0.2$ & $0.480\pm0.035$ & $3.170\pm0.231$ & $0.214\pm0.014$ & $0.049\pm0.004$\\
& $-0.2 \leq \log(L/L_{\rm Edd}) < 0.0$ & $0.564\pm0.042$ & $4.084\pm0.299$ & $0.252\pm0.017$ & $0.049\pm0.004$\\
\noalign{\smallskip}
$8.8 \leq \log(M_{\rm BH}/M_\odot) < 9.0$ & $-0.8 \leq \log(L/L_{\rm Edd}) < -0.6$ & $0.324\pm0.025$ & $2.545\pm0.190$ & $0.240\pm0.018$ & $0.053\pm0.004$\\
& $-0.6 \leq \log(L/L_{\rm Edd}) < -0.4$ & $0.512\pm0.037$ & $2.902\pm0.209$ & $0.255\pm0.020$ & $0.056\pm0.004$\\
& $-0.4 \leq \log(L/L_{\rm Edd}) < -0.2$ & $0.661\pm0.050$ & $4.920\pm0.361$ & $0.278\pm0.019$ & $0.060\pm0.005$\\
& $-0.2 \leq \log(L/L_{\rm Edd}) < 0.0$ & $0.647\pm0.049$ & $4.978\pm0.365$ & $0.262\pm0.019$ & $0.051\pm0.004$\\
\noalign{\smallskip}
$9.0 \leq \log(M_{\rm BH}/M_\odot) < 9.2$ & $-1.0 \leq \log(L/L_{\rm Edd}) < -0.8$ & $0.382\pm0.030$ & $3.268\pm0.246$ & $0.352\pm0.027$ & $0.016\pm0.001$\\
& $-0.8 \leq \log(L/L_{\rm Edd}) < -0.6$ & $0.577\pm0.045$ & $3.623\pm0.270$ & $0.334\pm0.026$ & $0.060\pm0.005$\\
& $-0.6 \leq \log(L/L_{\rm Edd}) < -0.4$ & $0.668\pm0.052$ & $4.429\pm0.329$ & $0.309\pm0.023$ & $0.058\pm0.005$\\
& $-0.4 \leq \log(L/L_{\rm Edd}) < -0.2$ & $0.652\pm0.050$ & $4.685\pm0.348$ & $0.300\pm0.024$ & $0.065\pm0.005$\\
& $-0.2 \leq \log(L/L_{\rm Edd}) < 0.0$ & $0.808\pm0.061$ & $5.783\pm0.427$ & $0.307\pm0.023$ & $0.071\pm0.006$\\
\noalign{\smallskip}
$9.2 \leq \log(M_{\rm BH}/M_\odot) < 9.4$ & $-1.0 \leq \log(L/L_{\rm Edd}) < -0.8$ & $0.731\pm0.058$ & $4.802\pm0.359$ & $0.363\pm0.026$ & $0.073\pm0.006$\\
& $-0.8 \leq \log(L/L_{\rm Edd}) < -0.6$ & $0.806\pm0.063$ & $4.841\pm0.362$ & $0.410\pm0.033$ & $0.104\pm0.008$\\
& $-0.6 \leq \log(L/L_{\rm Edd}) < -0.4$ & $0.805\pm0.063$ & $5.147\pm0.385$ & $0.384\pm0.033$ & $0.086\pm0.007$\\
& $-0.4 \leq \log(L/L_{\rm Edd}) < -0.2$ & $0.827\pm0.065$ & $5.618\pm0.420$ & $0.359\pm0.030$ & $0.086\pm0.007$\\
& $-0.2 \leq \log(L/L_{\rm Edd}) < 0.0$ & $1.026\pm0.078$ & $6.317\pm0.464$ & $0.335\pm0.026$ & $0.081\pm0.006$\\
\noalign{\smallskip}
$9.4 \leq \log(M_{\rm BH}/M_\odot) < 9.6$ & $-1.2 \leq \log(L/L_{\rm Edd}) < -1.0$ & $0.533\pm0.044$ & $3.877\pm0.298$ & $0.485\pm0.038$ & -- $^b$\\
& $-1.0 \leq \log(L/L_{\rm Edd}) < -0.8$ & $0.722\pm0.059$ & $4.545\pm0.346$ & $0.414\pm0.031$ & $0.089\pm0.007$\\
& $-0.8 \leq \log(L/L_{\rm Edd}) < -0.6$ & $0.848\pm0.069$ & $5.193\pm0.395$ & $0.463\pm0.035$ & $0.079\pm0.006$\\
& $-0.6 \leq \log(L/L_{\rm Edd}) < -0.4$ & $0.971\pm0.077$ & $6.341\pm0.478$ & $0.444\pm0.040$ & $0.092\pm0.007$\\
& $-0.4 \leq \log(L/L_{\rm Edd}) < -0.2$ & $1.036\pm0.082$ & $7.259\pm0.545$ & $0.442\pm0.036$ & $0.101\pm0.008$\\
\noalign{\smallskip}
$9.6 \leq \log(M_{\rm BH}/M_\odot) < 9.8$ & $-1.2 \leq \log(L/L_{\rm Edd}) < -1.0$ & $0.682\pm0.057$ & $3.951\pm0.305$ & $0.450\pm0.033$ & $0.021\pm0.002$\\
& $-1.0 \leq \log(L/L_{\rm Edd}) < -0.8$ & $0.969\pm0.080$ & $5.304\pm0.407$ & $0.511\pm0.045$ & $0.080\pm0.007$\\
& $-0.8 \leq \log(L/L_{\rm Edd}) < -0.6$ & $0.923\pm0.076$ & $5.463\pm0.419$ & $0.521\pm0.040$ & $0.093\pm0.008$\\
& $-0.6 \leq \log(L/L_{\rm Edd}) < -0.4$ & $0.971\pm0.079$ & $6.509\pm0.498$ & $0.496\pm0.045$ & $0.097\pm0.008$\\
& $-0.4 \leq \log(L/L_{\rm Edd}) < -0.2$ & $1.007\pm0.080$ & $7.315\pm0.552$ & $0.423\pm0.034$ & $0.088\pm0.007$\\
\noalign{\smallskip}
$9.8 \leq \log(M_{\rm BH}/M_\odot) < 10.0$ & $-1.2 \leq \log(L/L_{\rm Edd}) < -1.0$ & $0.966\pm0.082$ & $4.500\pm0.351$ & $0.600\pm0.048$ & $0.138\pm0.011$\\
& $-1.0 \leq \log(L/L_{\rm Edd}) < -0.8$ & $0.968\pm0.082$ & $5.455\pm0.424$ & $0.597\pm0.047$ & $0.106\pm0.009$\\
& $-0.8 \leq \log(L/L_{\rm Edd}) < -0.6$ & $1.186\pm0.099$ & $6.732\pm0.518$ & $0.603\pm0.055$ & $0.108\pm0.009$\\
& $-0.6 \leq \log(L/L_{\rm Edd}) < -0.4$ & $1.459\pm0.118$ & $8.694\pm0.662$ & $0.586\pm0.051$ & $0.118\pm0.010$\\
\noalign{\smallskip} \hline
\end{tabular}

\begin{list}{}{}
\item[$^a$]
Line flux ratios of \ion{N}{v}$\lambda$1240/\ion{C}{iv}$\lambda$1549, \ion{N}{v}$\lambda$1240/\ion{He}{ii}$\lambda$1640, (\ion{Si}{iv}$\lambda$1398+\ion{O}{iv}]$\lambda$1402)/\ion{C}{iv}$\lambda$1549, and \ion{Al}{iii}$\lambda$1857/\ion{C}{iv}$\lambda$1549. These errors are estimated by taking into account both random error and systematic error (five percent on line flux) for each emission line.
\item[$^b$]
In this interval \ion{Al}{iii}$\lambda$1857 is not detected.
\end{list}

\end{table*}


\begin{figure*}[!]
\centering
\begin{tabular}{c}
\includegraphics[width=17.4cm]{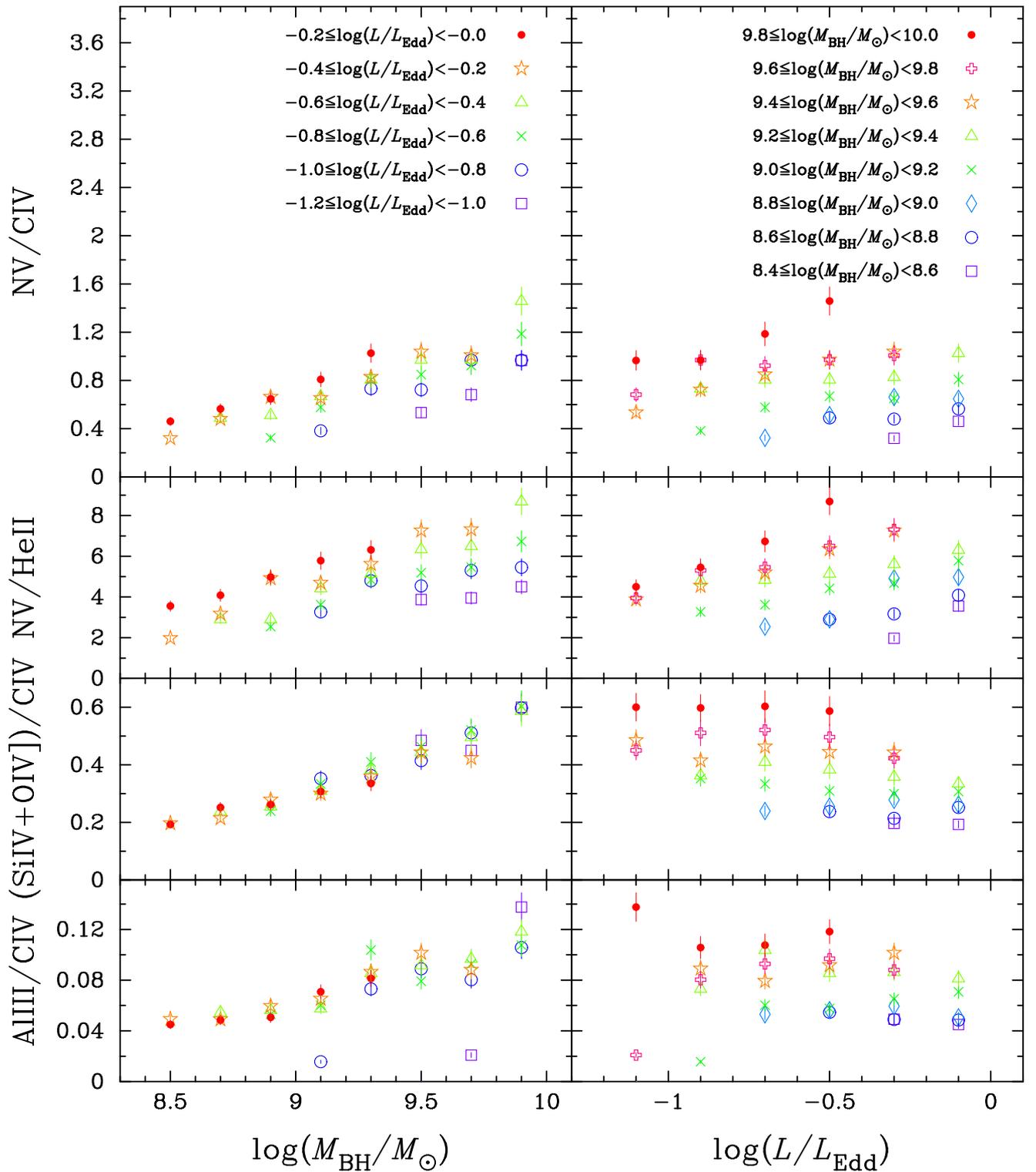}
\end{tabular}
\caption{Relation between the metallicity sensitive emission-line flux ratios and the black hole mass (left hand panels), and the Eddington ratio (right hand panels).}
\end{figure*}


\begin{figure}[!]
\centering
\begin{tabular}{c}
\includegraphics[width=8.5cm]{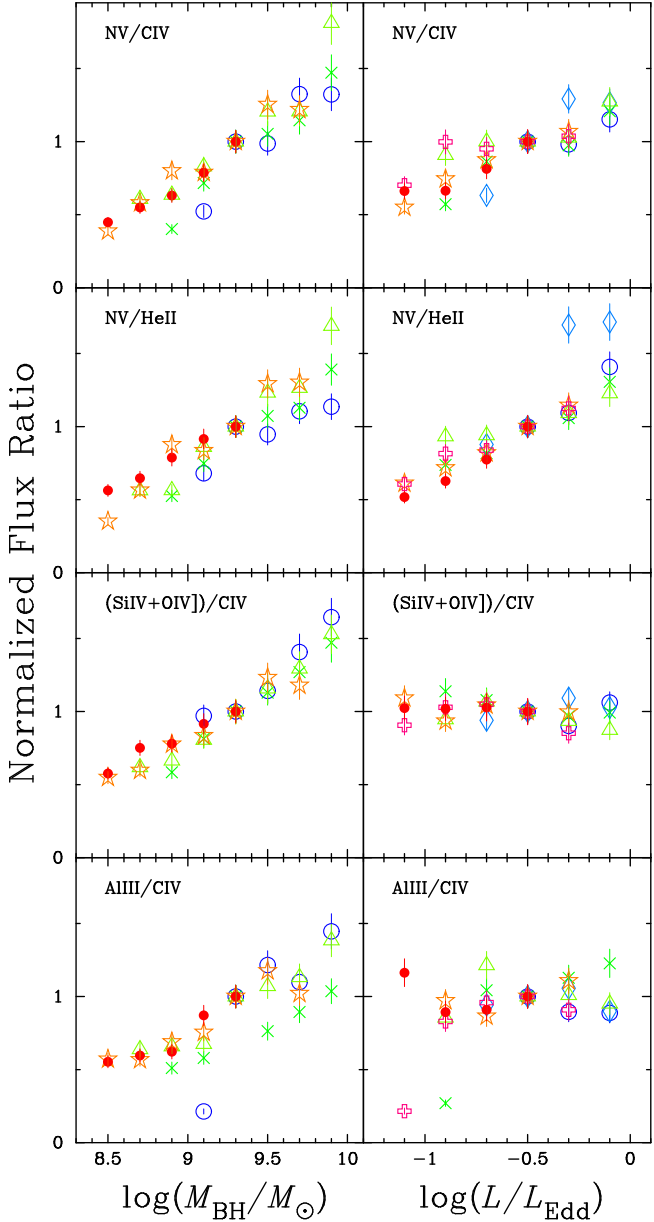}
\end{tabular}
\caption{Metallicity sensitive emission line flux ratios normalized by the value measured in the composite spectra with $-0.6 \leq \log (L/L_{\rm Edd}) < -0.4$ for the individual black hole mass bins, as a function of black hole mass (left hand panels). Emission line flux ratios normalized by the value at $9.2 \leq \log (M_{\rm BH}/M_\odot) < 9.4$ for the individual Eddington ratio bins, as a function of Eddington ratio (right hand panels). Symbols are the same as those in Figure 5.}
\end{figure}


\begin{table}
\caption{Results of the Spearman rank-order test.}
\begin{tabular}{l l l}
\hline \hline \noalign{\smallskip}
\multicolumn{1}{c}{line ratio} & \multicolumn{1}{c}{$r_{\rm s}$ $^a$} & \multicolumn{1}{c}{$p(r_{\rm s})$ $^b$}\\
\noalign{\smallskip} \hline \noalign{\smallskip}
\multicolumn{3}{c}{vs. black hole mass}\\
\noalign{\smallskip} \hline \noalign{\smallskip}
\ion{N}{v}/\ion{C}{iv} & $+0.94$ & $5.03\times10^{-15}$\\
\ion{N}{v}/\ion{He}{ii} & $+0.94$ & $9.08\times10^{-15}$\\
(\ion{Si}{iv}+\ion{O}{iv}])/\ion{C}{iv} & $+0.99$ & $2.20\times10^{-16}$\\
\ion{Al}{iii}/\ion{C}{iv} & $+0.87$ & $3.44\times10^{-10}$\\
\noalign{\smallskip} \hline \noalign{\smallskip}
\multicolumn{3}{c}{vs. Eddington ratio}\\
\noalign{\smallskip} \hline \noalign{\smallskip}
\ion{N}{v}/\ion{C}{iv} & $+0.86$ & $5.09\times10^{-10}$\\
\ion{N}{v}/\ion{He}{ii} & $+0.97$ & $2.20\times10^{-16}$\\
(\ion{Si}{iv}+\ion{O}{iv}])/\ion{C}{iv} & $-0.28$ & $1.24\times10^{-1}$\\
\ion{Al}{iii}/\ion{C}{iv} & $+0.38$ & $3.60\times10^{-2}$\\
\noalign{\smallskip} \hline
\end{tabular}

\begin{list}{}{}
\item[$^a$]
Spearman rank-order correlation coefficient.
\item[$^b$]
Probability of the data being consistent with the null hypothesis that the flux ratio is not correlated with the black hole mass or the Eddington ratio.
\end{list}

\end{table}


\section{Discussion}

\subsection{The black hole mass-metallicity relation}

As presented in section 4 (see also Figures 5 and 6), the tight relation between $M_{\rm BH}$ and {\it all} metallicity sensitive line ratios indicates that the $L_{\rm AGN}$-$Z_{\rm BLR}$ correlation is driven mostly by the $M_{\rm BH}$-$Z_{\rm BLR}$ relation (and not by any $L/L_{\rm Edd}$-$Z_{\rm BLR}$ relation). The $M_{\rm BH}$-$Z_{\rm BLR}$ relation can be understood in the following terms. If the $M_{\rm BH}$-$M_{\rm bul}$ relation observed in the local universe (Magorrian et al. 1998) holds at $z \sim 2 - 3$ (Janke et al. 2009), the $M_{\rm BH}$-$Z_{\rm BLR}$ relation obtained in this study implies a $M_{\rm bul}$-$Z_{\rm BLR}$ relation. In addition, if the BLR metallicity is associated with the metallicity of the host galaxy, the $M_{\rm BH}$-$Z_{\rm BLR}$ relation would simply trace the well-known correlation between the mass and metallicity in galaxies, i.e., the $M_{\rm bul}$-$Z_{\rm bul}$ relation (Lequeux et al. 1979; Tremonti et al. 2004; Maiolino et al. 2008). This implies that the origin of the $L_{\rm AGN}$-$Z_{\rm BLR}$ relation is the $M_{\rm bul}$-$Z_{\rm bul}$ relation (or, more simply, the ``mass-metallicity relation'').

The one potential problem of this interpretation is the lack of redshift evolution of the $L_{\rm AGN}$-$Z_{\rm BLR}$, out to $z \sim 6$ (Nagao et al. 2006a; Juarez et al. 2009; see also Dietrich et al. 2003), which is at odds with the redshift evolution of the mass-metallicity relation observed in galaxies (Maiolino et al. 2008, Mannucci et al. 2009). There are two possible solution to this problem. More recently, Mannucci et al. (2010) have found that the mass-metallicity relation in galaxies also depends on the SFR, more active galaxies being characterized by lower metallicity. They also found that, such mass-metallicity-SFR relation does not evolve with redshift out to $z \sim 2.5$; previous claims on the evolution of the mass-metallicity relation were simply a consequence of the fact that high-$z$ surveys have probed galaxies on average with higher SFR at high redshift. If quasar host galaxies at different redshifts are characterized by similar SFR (e.g., Maiolino et al. 2007a; Netzer et al. 2007; Lutz et al. 2008), this would imply that they are likely characterized by the mass-metallicity relation. However, this explanation is not totally satisfactory, since actually there is evidence that even the mass-metallicity-SFR relation evolves at $z > 3$ (Mannucci et al. 2010), while the $L_{\rm AGN}$-$Z_{\rm BLR}$ in quasars does not evolve out to $z \sim 6$.

Another effect, which goes in the direction of making the non-evolution of the $L_{\rm AGN}$-$Z_{\rm BLR}$ relation even more problematic, is that at high redshift the $M_{\rm BH}$-$M_{\rm bul}$ relation seems to evolve towards higher $M_{\rm BH}/M_{\rm bul}$ relative to the local relation (Walter et al. 2004; Woo et al. 2006, 2008; Maiolino et al. 2007b; Lamastra et al. 2010; Merloni et al. 2010). As a consequence, for a given black hole mass (hence similar luminosities, for a given accretion rate) high redshift AGNs should be associated with less massive, hence lower metallicity, host galaxies relative to AGNs at lower redshift.

A possible redshift evolution of the accretion rate $L/L_{\rm Edd}$ (at a given black hole mass, Netzer \& Trakhtenbrot 2007), would make the problem of the non evolution of the $L_{\rm AGN}$-$Z_{\rm BLR}$ relation even more serious. Indeed, at a given luminosity a high-$z$ AGN would have a black hole mass lower than a lower redshift AGN, hence should be associated with a less massive, and less enriched, host galaxy.

A possible explanation is that luminous quasars host galaxies do not follow the same secular evolutionary process as most of the galaxies used to investigate the metallicity evolution, but are instead characterized by vigorous star formation likely associated with merging events (e.g., Shao et al. 2010), yielding a very rapid co-evolution of the black hole and its host galaxy, implying a very rapid metals enrichment. Within such strong an rapid co-evolutionary scenario it is likely that the metallicity investigation of quasars host galaxies through the cosmic epochs is subject to strong selection effects. More specifically, as discussed in detail in Juarez et al. (2009), because of the co-evolutionary link between black holes and their host galaxies, quasars are detectable in flux-limited surveys only once their host galaxies are already evolved and chemically enriched, and this connection is the same at any epoch, regardless of redshift (see also Kawakatu et al. 2003).

\subsection{The dependence between the Eddington ratio and emission-line ratios involving \ion{N}{v}}

In section 4, we have also found that the emission-line flux ratios involving \ion{N}{v} are related to $L/L_{\rm Edd}$ whereas the other two emission-line flux ratios, i.e., (\ion{Si}{iv}+\ion{O}{iv}])/\ion{C}{iv} and \ion{Al}{iii}/\ion{C}{iv}, do not show any dependence on $L/L_{\rm Edd}$. If the relation between the Eddington ratio and emission-line flux ratios involving \ion{N}{v} is ascribed to a $L/L_{\rm Edd}$-$Z_{\rm BLR}$ relation, then the other two emission-line flux ratios should also show dependence on $L/L_{\rm Edd}$, since all four emission-line flux ratios are sensitive to the gas metallicity of the BLR. These results thus indicate that emission-line flux ratios involving \ion{N}{v} do not simply depend on metallicity but also on other independent parameters. One possible idea is that the difference of the behavior between flux ratios with and without \ion{N}{v} may be associated with the different enrichment timescale of nitrogen relative to other elements.

Recent studies on local AGNs have shown that nuclear star formation and AGN activity are not coeval (Davies et al. 2007; see also Wild et al. 2010). More specifically, black hole accretion appears to occur in post-starburst nuclei, with a delay of $\sim 10^8$ years from the starburst. Probably during the active starburst phase powerful supernova explosions expel the circumnuclear gas preventing it to reach the black hole, while the more gentle winds of AGB stars occurring on timescales of $10^8$ years are capable of stirring the interstellar matter (ISM), making it loose angular momentum and then feed the AGN. The production of nitrogen is delayed with respect to other elements, and indeed produced mostly by AGB stars whereas other elements (i.e., Si, O, and Al) are produced by massive stars on much shorter timescale. Carbon originates mainly from low-mass stars (Chiappini et al. 2003); their life time is $\sim 10^{9-10}$ years. This means that carbon would not be associated with black hole accretion. Moreover, since \ion{C}{iv}$\lambda$1549 is a strong coolant of the BLR, its intensity is little sensitive to the carbon abundance. As a consequence, if also at high-$z$ black hole accretion is associated with the delayed onset of AGB stars, one naturally expects a correlation between Eddington ratio and nitrogen abundance, as observed by us. Note that, in this picture, the correlation between Eddington ratio and emission-line flux ratios involving \ion{N}{v} suggest that the timescale of AGN feeding are shorter than the timescale of the nitrogen enrichment by AGB stars ($< 10^8$ years).

The relation between $L/L_{\rm Edd}$ and nitrogen abundance is associated to black hole accretion and star formation, i.e. on phenomena occurring on relatively short time scales ($\sim 10^8$ years). Conversely, the relation between black hole mass and metallicity is more fundamental and representative of the global evolution of these systems, since it connects physical quantities integrated over the whole formation history of black holes and galaxies.

Within this context it is interesting to note that NLS1s, which are characterized by very high Eddington ratios, have high \ion{N}{v}/\ion{C}{iv} ratio (Shemmer et al. 2004), in line with the trend found above for luminous high-$z$ quasars and suggesting that NLS1s have undergone vigorous star formation (Nagao et al. 2002).

It should be noted that previous studies have found that quasar hosts are characterized by vigorous {\it ongoing} star formation (e.g., Maiolino et al. 2007a; Netzer et al. 2007; Lutz et al. 2008). Our finding of a correlation between black hole accretion rate and nitrogen abundance suggests that at earlier times, by about 10$^8$ years, star formation was even higher.


\section{Conclusions}

We have investigated the relationship between the metallicity and SMBH mass or Eddington ratio by producing composite SDSS quasar spectra with 33 subsamples divided in intervals of black hole mass and Eddington ratio at $2.3 < z < 3.0$. In each of these spectra we measure emission-line flux ratios that are sensitive to the metallicity of the BLR gas. We have investigated the $M_{\rm BH}$-$Z_{\rm BLR}$ dependence or the $L/L_{\rm Edd}$-$Z_{\rm BLR}$ dependence by comparing the emission-line flux ratios of the composite spectra by fixing the Eddington ratio or black hole mass, respectively. We have found the following results:
\begin{enumerate}
\item
There is a tight correlation between the BLR metallicity and black hole mass, while the BLR metallicity is almost independent of the Eddington ratio, implying that the $L_{\rm AGN}$-$Z_{\rm BLR}$ relation in quasars results from the relationship between $Z_{\rm BLR}$ and $M_{\rm BH}$ (and not from a relationship between $Z_{\rm BLR}$ and $L/L_{\rm Edd}$).
\item
The emission-line flux ratios involving \ion{N}{v} (i.e., \ion{N}{v}/\ion{C}{iv} and \ion{N}{v}/\ion{He}{ii}) correlate with the Eddington ratio (while the other two metallicity sensitive emission-line flux ratios do not show any correlation with $L/L_{\rm Edd}$). This result suggests that the accretion rate in quasars is associated with a post-starburst phase, when AGB stars enrich the interstellar medium with nitrogen.
\end{enumerate}
By assuming that the $M_{\rm BH}$-$M_{\rm bul}$ relation applies at high-redshift, the $M_{\rm BH}$-$Z_{\rm BLR}$ relation is likely a consequence of the $M_{\rm bul}$-$Z_{\rm bul}$ relation in galaxies.

The lack of redshift evolution of the $L_{\rm AGN}$-$Z_{\rm BLR}$ relation does not necessarily imply a lack of metallicity evolution in AGN host galaxies. Indeed, as already suggested in previous works (e.g., Juarez et al. 2009), a combination of selection effects and of the co-evolution between black hole and galaxies causes quasars to be detected only once their host galaxy is already chemically evolved at any epoch, regardless of redshift.

\begin{acknowledgements}

We would like to thank M. Dietrich for his useful suggestions and comments. K.M. acknowledges financial support from the Japan Society for the Promotion of Science (JSPS) through JSPS Research Fellowships for Young Scientists, and form the Circle for the Promotion of Science and Engineering of Japan. T.N. acknowledges financial supports through the Research Promotion Award of Ehime University, the Kurata Grant from the Kurata Memorial Hitachi Science and Technology Foundation, and the Research Grant from the Itoh Science Foundation. R.M. and A.M. acknowledge partial support from INAF and from ASI through contract ASI-INAF I/016/07/0. Y.T. is financially supported by JSPS (Grant Nos. 17253001 and 19340046). Data analysis were in part carried out on common use data analysis computer system at the Astronomy Data Center, ADC, of the National Astronomical Observatory of Japan. Funding for the SDSS and SDSS-II has been provided by the Alfred P. Sloan Foundation, the Participating Institutions, the National Science Foundation, the U.S. Department of Energy, the National Aeronautics and Space Administration, the Japanese Monbukagakusho, the Max Planck Society, and the Higher Education Funding Council for England. The SDSS Web Site is http://www.sdss.org/. The SDSS is managed by the Astrophysical Research Consortium for the Participating Institutions. The Participating Institutions are the American Museum of Natural History, Astrophysical Institute Potsdam, University of Basel, University of Cambridge, Case Western Reserve University, University of Chicago, Drexel University, Fermilab, the Institute for Advanced Study, the Japan Participation Group, Johns Hopkins University, the Joint Institute for Nuclear Astrophysics, the Kavli Institute for Particle Astrophysics and Cosmology, the Korean Scientist Group, the Chinese Academy of Sciences (LAMOST), Los Alamos National Laboratory, the Max-Planck-Institute for Astronomy (MPIA), the Max-Planck-Institute for Astrophysics (MPA), New Mexico State University, Ohio State University, University of Pittsburgh, University of Portsmouth, Princeton University, the United States Naval Observatory, and the University of Washington.

\end{acknowledgements}


\end{document}